\documentclass[twocolumn,aps,prd,floatfix,preprintnumbers,a4paper,nofootinbib,superscriptaddress]{revtex4-1}
 
\usepackage{graphicx}				
\usepackage{placeins}
\usepackage{amssymb}
\usepackage{amsmath}
\usepackage[dvipsnames]{xcolor}
\usepackage{verbatim}
\usepackage{rotating}
\usepackage{comment}

\usepackage{graphicx}
\usepackage{amsfonts}
\usepackage{dcolumn}
\usepackage{longtable}
\usepackage[caption=false]{subfig}
\usepackage{soul}
\usepackage{mathrsfs}
\usepackage{enumitem} 
\usepackage{tabularx}
\usepackage{subfig}
\usepackage{soul}

\usepackage{natbib}
\usepackage[colorlinks]{hyperref}

\definecolor{LinkColor}{rgb}{0, 0, 0.75}
\definecolor{CiteColor}{rgb}{0.75, 0, 0}
\definecolor{UrlColor}{rgb}{0, 0, 0.75}
\hypersetup{linkcolor=LinkColor}
\hypersetup{citecolor=CiteColor}
\hypersetup{urlcolor=UrlColor}

\begin{document} 


\newenvironment{aside}
    {\begin{addmargin}[1em]{2em}
\begin{center} \noindent\rule{0.65\paperwidth}{0.4pt} \end{center}
    }
    {
    \begin{center} \noindent\rule{0.65\paperwidth}{0.4pt} \end{center}
\end{addmargin}
    }

\newcommand\xquote[1]{``#1"}
\newcommand\psinr[2]{$\psi_{{#1}{#2}}^{NR}$}
\newcommand\red[1]{{\color[rgb]{0.75,0.0,0.0} #1}}
\newcommand\redoff[1]{#1}
\newcommand\greenoff[1]{#1}
\newcommand\bred[1]{{\color[rgb]{0.75,0.0,0.0} \textbf{#1}}}
\newcommand\green[1]{{\color[rgb]{0.0,0.60,0.08} #1}}
\newcommand\blue[1]{{\color[rgb]{0,0.20,0.65} #1}}
\newcommand\cyan[1]{{\color[HTML]{00c3ff} #1}}
\newcommand\bluey[1]{{\color[rgb]{0.11,0.20,0.4} #1}}
\newcommand\gray[1]{{\color[rgb]{0.7,0.70,0.7} #1}}
\newcommand\grey[1]{{\color[rgb]{0.7,0.70,0.7} #1}}
\newcommand\white[1]{{\color[rgb]{1,1,1} #1}}
\newcommand\darkgray[1]{{\color[rgb]{0.3,0.30,0.3} #1}}
\newcommand\orange[1]{{\color[rgb]{.86,0.24,0.08} #1}}
\newcommand\purple[1]{{\color[rgb]{0.45,0.10,0.45} #1}}
\newcommand\note[1]{\colorbox[rgb]{0.85,0.94,1}{\textcolor{black}{\textsc{\textsf{#1}}}}}

\newcommand*{\figfactor}{0.495}

\def\m#1{{\hat{#1}}} 
\def\ma#1{{\hat{#1}_{\bm{a}}}} 
\newcommand{\var}[1]{\mathcal{{#1}}}
\newcommand{\mlam}{{\bm{\lambda}}}
\newcommand{\lam}{{\lambda}}
\newcommand{\mLam}{{\bm{\Lambda}}}
\newcommand{\bigo}[1]{{\cal O}({#1})}
\newcommand{\braket}[2]{ {\langle {#1} \, | \, {#2} \rangle} }
\newcommand{\bra}[1]{ \langle {#1} |  }
\newcommand{\ket}[1]{ | {#1} \rangle }
\newcommand{\ketbra}[2]{ \ket{#1}\bra{#2} }
\newcommand{\tx}[1]{\text{#1}}
\def\T{\dagger}

\definecolor{lightblue}{rgb}{.82,.88,0.95}
\definecolor{lightred}{rgb}{0.95,.86,0.86}
\definecolor{yellow}{rgb}{0.95,0.95,0.86}
\definecolor{green}{rgb}{.90,1,0.95}
\definecolor{lightpurple}{rgb}{.95,0.85,0.95}

\def\prd{Phys.Rev.D}
\def\gt{Georgia Tech}
\def\tk{{Teukolsky}}
\def\mpl{multipole}
\def\ee{Einstein's equations}
\def\toolkit#1{NRDA--Toolkit{#1}}
\def\wf#1{waveform#1}
\def\gr#1{General Relativity#1
  (GR#1)\gdef\gr{GR}}
\def\gwa#1{Gravitational Wave Astrophysics#1}
\def\gwf#1{gravitational waveform#1}
\def\gwa#1{\gw{} astronomy#1}
\def\grad#1{gravitational radiation#1}
\def\nht#1{No-Hair Theorem#1}
\def\rm#1{\mathrm{#1}}
\def\BL{Boyer-Lindquist}

\def\prt#1{Part~(\ref{#1})}
\def\Apx#1{Appendix~(\ref{#1})}
\def\apx#1{Appx.~(\ref{#1})}
\def\capx#1{Appx.~\ref{#1}}
\def\ch#1{Chapter~(\ref{#1})}
\def\cch#1{Chapter~\ref{#1}}
\newcommand{\chs}[2]{Chapters~(\ref{#1}-\ref{#2})}
\newcommand{\cchs}[2]{Chapters~\ref{#1}-\ref{#2}}
\def\Sec#1{Section~\ref{#1}}
\def\sec#1{Sec.~\ref{#1}}
\def\csec#1{Sec.~\ref{#1}}
\newcommand{\csecs}[2]{Secs.~\ref{#1}-\ref{#2}}
\def\tk#1{Teukolsky#1}
\newcommand{\secs}[2]{Secs.~\ref{#1}-\ref{#2}}
\newcommand{\secsa}[2]{Sec.~\ref{#1} and Sec.~\ref{#2}}
\def\Tbl#1{Table~(\ref{#1})}
\def\tbl#1{Table~(\ref{#1})}
\def\ctbl#1{Table~\ref{#1}}
\def\Fig#1{Figure~\ref{#1}}
\def\fig#1{Fig.~\ref{#1}}
\def\cfig#1{Fig.~\ref{#1}}
\newcommand{\figs}[2]{Figures~(\ref{#1}-\ref{#2})}
\newcommand{\Figs}[2]{Figures~(\ref{#1}-\ref{#2})}
\newcommand{\Figsa}[2]{Figures~(\ref{#1}) and (\ref{#2})}
\def\Eqn#1{Equation~(\ref{#1})}
\def\eqn#1{Eq.~(\ref{#1})}
\def\ceqn#1{Eq.~\ref{#1}}
\newcommand{\Eqns}[2]{Equations~(\ref{#1}-\ref{#2})}
\newcommand{\Eqnsa}[2]{Equations~(\ref{#1}) and (\ref{#2})}
\newcommand{\eqns}[2]{Eqs.~(\ref{#1}-\ref{#2})}
\newcommand{\eqnsa}[2]{Eqs.~(\ref{#1}) and (\ref{#2})}
\newcommand{\ceqns}[2]{Eqs.~\ref{#1}-\ref{#2}}
\newcommand{\ceqnsa}[2]{Eqs.~\ref{#1} and \ref{#2}}

\def\tridiag#1{tridiagonal#1}
\def\ply#1{{polynomial#1}}
\def\plys{{polynomials}}
\def\orthog{orthogonality}
\def\da#1{data analysis#1}
\def\sw{spin weighted}
\def\swsh{spin weighted spherical harmonic}
\def\lal#1{LIGO Analysis Library#1
  (LAL#1)\gdef\lal{LAL}}
\def\nrda#1{\nr{} Data Analysis#1
  (NRDA#1)\gdef\nrda{NRDA}}
\def\tt#1{\textit{transverse--traceless}#1
  (TT#1)\gdef\tt{TT}}
\def\et#1{Einstein Telescope#1
  (ET#1)\gdef\et{ET}}
\def\ce#1{Cosmic Explorer#1
  (CE#1)\gdef\ce{CE}}
\def\ego#1{European Gravitational Observatory#1
  (EGO#1)\gdef\ego{EGO}}
\def\lisa#1{Laser Interferometer Space Antenna#1
  (LISA#1)\gdef\lisa{LISA}}
\def\ligo#1{Laser Interferometer Gravitational Wave Observatory#1
  (LIGO#1)\gdef\ligo{LIGO}}
\def\igwn#1{International Gravitational Wave Network#1
  (IGWN)\gdef\igwn{IGWN}}
\def\lvk#1{LIGO-Virgo-Kagra collaboration#1
  (LVK)\gdef\lvk{LVK}}
\def\lv#1{#1
(LIGO-Virgo#1)\gdef\lv{LV}}
\def\virgo#1{Virgo#1}
\def\aligo#1{Advanced LIGO#1
  (Adv. LIGO#1)\gdef\aligo{Adv. LIGO}}
\def\snr#1{signal to noise ratio#1
  (SNR#1)\gdef\snr{SNR}}
\def\psd#1{power spectral density#1
  (PSD#1)\gdef\psd{PSD}}
\def\rom#1{reduced order model#1
  (ROM#1)\gdef\rom{ROM}}
\def\gatech#1{Georgia Institute of Technology#1
  (GaTech#1)\gdef\gatech{GaTech}}
\def\ffi#1{Fixed-Frequency Integration#1
  (FFI#1)\gdef\ffi{FFI}}
\def\sxs#1{Simulating Extreme Spacetimes#1
  (SXS#1)\gdef\sxs{SXS}}
\def\bam#1{Bifunctional Adaptive Mesh#1
  (BAM#1)\gdef\bam{BAM}}
\def\adm#1{Arnowitt-Deser-Misner
	(ADM#1)\gdef\adm{ADM}}
\def\frmse#1{Fractional Root-Mean Square Error
	(FRMSE)\gdef\frmse{FRMSE}}

\def\bh#1{black hole#1
 (BH#1)\gdef\bh{BH}}
\def\bbh#1{binary black hole#1
 (BBH#1)\gdef\bbh{BBH}}
\def\bhb#1{\bh{} binary#1}

\def\qnm#1{Quasi-Normal Mode#1
(QNM#1)\gdef\qnm{QNM}}
\def\Qnm#1{Quasi-Normal Mode#1}
\def\Qnms{Quasi-Normal Modes}
\def\eob#1{Effective One Body#1
  (EOB#1)\gdef\eob{EOB}}
\def\sws#1{spheroidal harmonics of spin weight -2#1}
\def\swy#1{spherical harmonics of spin weight -2#1}
\def\gw#1{gravitational wave#1}
\def\gwa#1{gravitational wave astronomy#1}
\def\pn#1{Post-Newtonian#1
 (PN#1)\gdef\pn{PN}}
\def\pnl#1{post-Newtonian-like#1
  (PN-like#1)\gdef\pnl{PN-like}}
\def\NR{{\text{NR}}}
\def\nr{Numerical Relativity
 (NR)\gdef\nr{NR}}
\def\pt{\bh{} perturbation theory}
\def\GOLS#1{\textit{greedy ordinary least-squares}#1}
\def\rd{ringdown}
\def\imr{inspiral-merger-ringdown}
\def\cbc#1{compact object coalescence#1}
\def\bbc#1{\bbh{} coalescence#1}
\def\pc#1{principle component#1}
\def\pca#1{principle component analysis#1
  (PCA#1)\gdef\pca{PCA}}
\def\svd#1{Singular Value Decomposition#1
  (SVD#1)\gdef\svd{SVD}}
\def\gs#1{Gram-Schmidt#1}
\newcommand{\y}[2]{ {_{#1}}Y_{#2} }
\def\sylm{ \y{s}{\ell m } }
\def\mo{\mathcal{D}}
\def\adjmo{{\mathcal{D}}^{\dagger}}
\def\yo{\mathcal{D}_Y}
\def\adjyo{{\mathcal{D}_Y}^{\hspace{-2pt}\dagger}}
\def\ro{\mathcal{D}_R}
\def\adjro{{\mathcal{D}_R}^{\hspace{-2pt}\dagger}}
\def\bo{\mathcal{D}_B}
\def\adjbo{{\mathcal{D}_B}^{\hspace{-2pt}\dagger}}
\def\so{\mathcal{D}_S}
\def\adjso{{\mathcal{D}_S}^{\hspace{-2pt}\dagger}}
\def\to{\mathcal{D}_T}
\def\adjto{{\mathcal{D}_T}^{\hspace{-2pt}\dagger}}
\def\adj#1{{#1}^{\dagger}}
\def\dadj#1{{#1}^{\ddag}}
\def\ethp{\eth}
\def\ethm{{\eth'}}
\def\A{{\L m}}
\def\a{{\alpha}}
\def\mcl{\mathcal{L}}
\def\mct{{\mathcal{T}}}
\def\mcv{{\mathcal{V}}}
\def\cmcv{{\mcv^*}}
\def\cmct{{\mct^*}}
\def\tmct{{\tilde{\mct}}}
\def\tmcv{{\tilde{\mcv}}}
\def\mclo{{\mathcal{L}_o}}
\def\mcto{{\mathcal{T}_o}}
\def\mcvo{{\mathcal{V}_o}}
\def\cmcvo{{{\mcv_o^*}}}
\def\cmcto{{{\mct_o^*}}}
\def\tmcto{{\tilde{\mct}_o}}
\def\tmcvo{{\tilde{\mcv}_o}}
\def\tmcl{\tilde{\mcl}}
\def\mcp{\mathcal{P}}
\def\mcq{\mathcal{Q}}
\def\amcl{\adj{\mcl}}
\def\cmcl{{\mcl^*}}
\def\sjk{{\sigma_{\lp\ell}}}
\def\mcD{\mathcal{D}}
\def\mcL{\mathcal{L}}
\def\mcT{\mathcal{T}}
\def\mcV{\mathcal{V}}
\def\mcP{\mathcal{P}}
\def\mcQ{\mathcal{Q}}
\def\Lo{ {\mathcal{K}} }
\def\I{{\mathbb{I}}}
\def\max{\mathrm{max}}

\def\gs{Gram-Schmidt}
\def\polys{polynomials}

\def\lMn{{{\ell \M n}}}
\def\lmn{{{\ell m n}}}
\def\lmin{{{\ell_\mathrm{min}}}}
\def\lpmn{{{\ell' m n}}}
\def\lpmnp{{{\ell' m n'}}}
\def\lm{{{\ell m}}}
\def\lpm{{{\ell' m}}}
\def\l{{{\ell}}}
\def\n{{\bar{n}}}
\def\lmbn{{{\ell m \n}}}
\def\lpmbn{{{\ell' m \n}}}
\def\lp{{{\ell'}}}
\def\pp{p}

\mathchardef\minus = "002D
\newcommand{\swY}[4][]{{}_{{}_{#2}}\!Y^{#1}_{#3}(#4)}
\newcommand{\swSH}[5][]{{}_{{}_{#2}}S^{#1}_{#3}(#4;#5)}
\newcommand{\swS}[5][]{{}_{{}_{#2}}S^{#1}_{#3}(#4;#5)}
\newcommand{\scA}[4][]{{}_{{}_{#2}}A^{#1}_{#3}(#4)}
\newcommand{\YSH}[3][]{\mathcal{A}^{#1}_{#2}(#3)}

\def\LMaster{ \mathcal{L}_{t r \theta \phi} }
\def\LMasterB{ \mathcal{L}_{t r u \phi} }
\def\rp{ r_{+} }
\def\rm{ r_{-} }

\def\rhs{right hand side}
\def\lhs{left hand side}
\def\te{\tk{'s} equation}

\def\L{\bar{\ell}}
\def\M{\bar{m}}
\def\LM{{\L\M}}
\def\Lop{\mathcal{L}_{\k}}

\newcommand{\brak}[2]{ \braket{#1}{#2} }
%
\newcommand*{\factor}{0.95} 
\newcommand*{\rscale}{1.3}

\newcommand{\hlgreen}[1]{\sethlcolor{green}\hl{#1}{\sethlcolor{yellow}}}
\newcommand{\hlyellow}[1]{\sethlcolor{yellow}\hl{#1}{\sethlcolor{yellow}}}
\newcommand{\hlblue}[1]{\sethlcolor{lightblue}\hl{#1}{\sethlcolor{yellow}}}
\newcommand{\hlred}[1]{\sethlcolor{lightred}\hl{#1}{\sethlcolor{yellow}}}
\newcommand{\hlpurple}[1]{\sethlcolor{lightpurple}\hl{#1}{\sethlcolor{yellow}}}

\newcommand{\qnms}{\qnm{s}}

\def\check#1{\red{#1}}
\def\new#1{\blue{#1}}
\def\changed#1{\underline{\textbf{\red{#1}}}}
\def\remove#1{\hlred{#1}}
\newcommand{\cw}{\tilde{\omega}}
\newcommand{\CW}{\tilde{\Omega}}
\newcommand{\CWr}{{\Omega}^{\mathrm{r}}}
\newcommand{\CWc}{{\Omega}^{\mathrm{c}}}
\newcommand{\SC}{\mathcal{K}}
\newcommand{\CC}{\mathcal{C}}
\newcommand{\SCr}{\mathcal{K}^{\mathrm{r}}}
\newcommand{\SCc}{\mathcal{K}^{\mathrm{c}}}
\newcommand{\lalapprox}{\texttt{MMRDNS}}
\def\jf{j_f}
\def\mf{M_f}
\newcommand{\LL}{\bar{l}}
\newcommand{\MM}{\bar{m}}
\def\gmvp#1{greedy-multivariate-polynomial#1
  (\texttt{GMVP}#1)\gdef\gmvp{\texttt{GMVP}}}
\def\gmvr#1{greedy-multivariate-rational#1
  (\texttt{GMVR}#1)\gdef\gmvr{\texttt{GMVR}}}

\def\PaperOne{\hyperlink{cite.London:202XP1}{Paper {I}}}
\def\PaperTwo{\hyperlink{cite.London:202XP2}{Paper {II}}}
\def\FigThreeA{Fig.~\hyperref[F3]{3a}}
\def\FigThreeB{Fig.~\hyperref[F3]{3b}}
\def\FigThreeC{Fig.~\hyperref[F3]{3c}}
\def\FigThreeD{Fig.~\hyperref[F3]{3d}}
\def\FigThreeE{Fig.~\hyperref[F3]{3e}}
\def\FigThreeF{Fig.~\hyperref[F3]{3f}}
\def\FigThreeG{Fig.~\hyperref[F3]{3g}}
\def\FigFourA{Fig.~\hyperref[F4]{4a}}
\def\FigFourB{Fig.~\hyperref[F4]{4b}}
\def\FigFourC{Fig.~\hyperref[F4]{4c}}
\def\ccHp#1{canonical confluent Heun polynomial#1}
\def\CcHp#1{Canonical confluent Heun polynomial#1}
\def\cHp#1{confluent Heun polynomial#1}
\def\CHp#1{Confluent Heun polynomial#1}
\def\cp#1{canonical polynomial#1}
\def\i{(\textit{i})}
\def\ii{(\textit{ii})}
\def\iii{(\textit{iii})}
\def\ci{\textit{i}}
\def\cii{\textit{ii}}
\def\ciii{\textit{iii}}
\def\monm#1{\langle\xi^{#1}\rangle}
\def\wrt{with respect to }
\def\pspm{{p_\star^\pm} }
\def\psp{{p_\star^+} }
\def\psm{{p_\star^-} }
\def\LcH{ \hat{L}{_{p}}^{(\tx{cH})} }
\def\CZero{{\tx{C}_0} }
\def\COne{{\tx{C}_1} }
\def\CTwo{{\tx{C}_2} }
\def\CThree{{\tx{C}_3} }
\def\CFour{{\tx{C}_4} }
\def\nnint{non-negative integer}

\def\bpar#1{\smallskip\smallskip\paragraph*{\textbf{#1}}--~}
\def\bparNoSkip#1{\paragraph*{\textbf{#1}}--~}
\def\iparNoSkip#1{\paragraph*{\textit{#1}}--~}
\def\tre{\tk{}'s radial equation}
\def\trp{the radial problem}

\newcommand{\KCL}{King's  College  London,  Strand,  London  WC2R  2LS,  United Kingdom}

\title{Natural polynomials for Kerr quasi-normal modes} 

\author{Lionel London} \affiliation{\KCL} 
\author{Michelle Foucoin} \affiliation{\KCL} 

\begin{abstract}
We present a polynomial basis that exactly tridiagonalizes Teukolsky's radial equation for quasi-normal modes. 
These polynomials naturally emerge from the radial problem, and they are ``canonical'' in that they possess key features of classical polynomials.
Our canonical polynomials may be constructed using various methods, the simplest of which is the Gram-Schmidt process. 
\redoff{In contrast with other polynomial bases, our polynomials allow for Teukolsky's radial equation to be represented as a simple matrix eigenvalue equation.}
We expect that our polynomials will be useful for better understanding the Kerr quasinormal modes' properties, particularly their prospective spatial completeness and orthogonality. 
We show that our polynomials are closely related to the confluent Heun and Pollaczek-Jacobi type polynomials.
Consequently, our construction of polynomials may be used to tridiagonalize other instances of the confluent Heun equation.
We apply our polynomials to a series of simple examples, including: (1) the high accuracy numerical computation of radial eigenvalues, (2) the evaluation and validation of quasinormal mode solutions to Teukolsky's radial equation, and (3) the use of Schwarzschild radial functions to represent those of Kerr.
Along the way, a potentially new concept, ``polynomial/non-polynomial duality'', is encountered and applied to show that some quasinormal mode separation constants are well approximated by confluent Heun polynomial eigenvalues.
We briefly discuss the implications of our results on various topics, including the prospective spatial completeness of Kerr quasinormal modes.
\end{abstract}
\maketitle
%
\section{Introduction}
\par The results of single \pt{} have proved widely useful to \gw{} astronomy~\cite{Vishv:1970,LIGOScientific:2016aoc,TheLIGOScientific:2016src,LIGOScientific:2020ibl,LIGOScientific:2021djp,LIGOScientific:2018mvr,Hughes:2019zmt,Khan:2015jqa,Husa:2015iqa,Pompili:2023tna}.
Most famously, the aftermath of \bbc{} is expected to be a perturbed Kerr \bh{} that rings down, radiating away its perturbation, predominantly in the form of damped oscillatory modes~\cite{Press:1971ApJ,Press:1973zz,Teukolsky:1973ha,TeuPre74_3,leaver85,Berti:2005ys,London:2014cma,Hamilton:2023znn}.
The related \textit{\rd{}} radiation is of practical relevance to many aspects of \gw{} astronomy, from the construction of signal models, to the ongoing search for new physics~\cite{London:2014cma,Hamilton:2023znn,Hughes:2019zmt,Khan:2015jqa,Husa:2015iqa,Pompili:2023tna,Carullo:2018sfu,Baibhav:2023clw,Carullo:2019flw,Cotesta:2022pci,Mishra:2010tp,Bhagwat:2019dtm,TheLIGOScientific:2016src}.
Concurrently, \rd{} radiation has inspired varied activity in \gw{} theory, from efforts to understand \bh{} nonlinearity, to emerging developments in \greenoff{the time-dependent perturbation theory of single \bh{s}}~\cite{Lagos:2022otp,Green:2022htq,Cannizzaro:2023jle,Sberna:2021eui,Redondo-Yuste:2023ipg}.
These efforts, among many others, aim to prepare \gw{} science for a future in which high-precision \rd{} measurements may point the way to a more complete understanding of gravity~\cite{LISA:2022kgy,Maselli:2017kvl,Berti:2016lat,Barausse:2014pra,Klein:2015hvg,Bailes2021,Vallisneri:2014vxa,Karnesis:2022vdp,LISA:2022kgy}. 
There is, therefore, good reason to think that a deeper practical understanding of linearly perturbed Kerr \bh{s} is of use. 
\par The present work is concerned with answering a number of questions regarding the mathematical structure and practical use of \gw{} \rd{}.
Ringdown is comprised of \qnm{s}, which are the analog of normal modes for dissipative systems~\cite{LivRevQNM,Nollert:1999ji,Berti:2009kk}.
Here, they will be treated under the \tk{} formalism, where the field equations are separable, and where the \qnm{s} are the joint point spectra of \tk{'s} \textit{radial} and \textit{angular} equations~\cite{Press:1973zz,Teukolsky:1973ha,TeuPre74_3,leaver85,Berti:2005ys,Cook:2014cta}.
In this context, the present work addresses the following questions about Kerr \qnm{s}.
\iparNoSkip{What is the nature of the overtone label?} Not long after the pioneering work of \tk{}, Press, and many others on linear perturbations of the Kerr geometry, Leaver lead the development of analytic methods for computing the \qnm{'s} functional dependence (i.e. in space and time) and mode numbers~\cite{leaver85,Leaver:1986JMP,Leaver86c}. 
Since, there has been tremendous progress in understanding the mathematical structure of Kerr \qnm{s}~\cite{Berti:2005ys,Berti:2003jh,Berti:2005gp,Berti:2006:ExFacs,Yang:2012pj,Yang:2012he,Zimmerman:2015rua,Zimmerman:2015trm,Maggiore:2007nq,Andersson:PhysRevD.51.353,Andersson:1996cm}.
For example, analysis of the gravitational field equations confers that \qnm{s} are described by four mode numbers, one for each spacetime dimension~\cite{Berti:2005ys,leaver85}.
In any asymptotically valid spacetime coordinates, these are $\cw_\lmn$ (time), $\ell$ (polar angle), $m$ (azimuthal angle), and $n$ (radius).
The polar mode number, $\ell$, is unavoidably linked to the eigenvalues of Jacobi polynomials~\cite{London:202XP1,Fackerell:1977,London:2020uva}.
In this sense, the Jacobi polynomials are \textit{natural} to the \qnm{} problem.  
Regarding the azimuthal angle $\phi$, this is similarly true of the monomials, $\zeta^m$, where $\zeta=e^{i\phi}$.
In contrast, the radial mode number, most commonly referred to as the ``overtone'' index\footnote{The use of ``overtone'' is known to be somewhat of a misnomer; e.g. acoustic overtones have higher frequencies compared to a fundamental resonant frequency. In contrast, \qnm{} overtones are known to have slightly lower frequencies (but higher damping rates) compared to the $n=0$, or fundamental, \qnm{s}.}, has evaded rudimentary explanation. 
\par Here, the overtone index is \greenoff{demonstrated to be closely related to} the order parameter inherent to the \cHp{s}~\cite{ronveaux1995heun,NIST:DLMF:ConfHeunPoly,Fiziev:2009ud,Hortacsu:2011rr,Dariescu:2021zve,MAGNUS2021105522}.
%
%
Concurrently, it is observed perhaps for the first time that the \cHp{s} exhibit a kind of ``polynomial/non-polynomial duality'', where finite sets of \cHp{s} are necessarily paired with an infinite set of non-polynomial functions.
\iparNoSkip{Is a simple spectral solution to \tk{'s} radial equation possible?} 
It was pointed out in Ref.~\cite{London:202XP1} (hereafter \PaperOne{}) that there exists a scalar product for which \tk{'s} radial equation\greenoff{, while non-Hermitian~\cite{ROSASORTIZ201826,Mostafazadeh:2001jk,Andrianov:2006xh,QNMTopicalReview09,Zhang:2020msa,Brody:2013ajot}}, is self-adjoint, and that this scalar product enables the direct application of Sturm-Liouville theory~\cite{pinchover_rubinstein_2005,abramowitz+stegun,ARFKEN2013401,Courant1954}.
The scalar product's weight function coincides with the Green's function approach to \qnm{s} pioneered by Leaver\footnote{\greenoff{The radial equation studied here is the same studied by Leaver~\cite{leaver85}. Forthcoming numerical results are ultimately equivalent to those from Leaver's series solution. However, truncations of Leaver's series solution (the infinite dimensional problem) should not be confused with the canonical polynomials presented in \sec{s4}, which are used to frame a finite dimensional problem in \sec{s5}.}}, and the deformed Jacobi weight known for Pollaczek{\textendash}Jacobi polynomials\cite{London:202XP1,Leaver86c,Chen:2010}.
One use of the scalar product is to represent \tk{}'s radial equation as a simple linear matrix equation~\cite{Axler:2015}. 
\par \greenoff{While many studies have developed matrix representations of \tk{'s} radial problem~(e.g.~\cite{Chung:2023zdq,Blazquez-Salcedo:2023hwg,Chung:2023wkd,ghojogh2023eigenvalue,Zhu:2023mzv}), the scalar product presented in \PaperOne{} was is perhaps the first to result in matrix representations that are explicitly symmetric and tridiagonal, meaning that all physically relevant information is encoded in the diagonal and upper-diagonal matrix elements.} 
Tridiagonal matrices are known to have number of favorable properties; e.g., \redoff{they are well developed methods to study their asymptotic structure~\cite{teschl2000jacobi,kuijlaars2003orthogonality,Chen:2010}, and the classical properties of related polynomials enable efficient numerical methods~\cite{Chen:2019,Chen:2010,chihara2011introduction}.} 
\greenoff{To arrive at a tridiagonal representation of \tk{}'s radial equation, it was postulated that there may exist a basis of problem-specific, or ``natural'', polynomials.}
\par Here, it is numerically shown that there do indeed exist unique classical-like polynomials that exactly tridiagonalize \tk{}'s radial equation. 
Therefore, the resulting method of solution to the \qnm{}'s radial problem is spectral in the simplest possible way: widely available linear algebra packages may be used to explicitly diagonalize the problem, simultaneously yielding eigenvalues and eigenfunctions.
The polynomials in question will be referred to as ``\ccHp{s}''; numerical examples are provided. 
\iparNoSkip{Is deeper analytic understanding possible?}
Leaver's algorithm for computing the \qnm{s'} functions and mode numbers has influenced many studies, particularly those of the \qnm{} frequencies, $\cw_\lmn$~\cite{Cook:2014cta,MaganaZertuche:2021syq,QNMTopicalReview09}.
Concurrently, methods such as the WKB and eikonal approximations have supported advances in the analytic understanding of \qnm{'s} behavior in certain limits, e.g. very high \bh{} spins, and/or very large values of $\ell$~\cite{Zimmerman:2015rua,Zimmerman:2015rua,Yang:2012he}.
Recently, there has been progress in understanding the pseudo-spectral stability of the \qnm{} frequency spectrum~\cite{Jaramillo:2020tuu,Boyanov:2023qqf,Destounis:2023ruj,Courty:2023rxk,Torres:2023nqg}.
Amid this activity, a number of related questions have remained open. These include:  
Do the \qnm{s'} radial dependence have an underlying structure that is polynomial?
Is it heuristically accurate to think of the radial mode number, $n$, as being the order of an underlying polynomial object, in the same way that $\ell$ are related to the order of a Jacobi polynomial?
\par Here, the \qnm{s'} radial functions are shown to be generally non-polynomial in nature, but potentially well approximated by a \cHp{} of order given by $n+1$. 
A simple procedure for quantifying the approximate polynomial nature of a \qnm{'s} radial function is presented.
\iparNoSkip{Are the \qnm{s} spatially complete?}
There has been a significant amount of research into the question of \qnm{} {completeness}\footnote{ A sequence of functions is complete (i.e. is a basis) if any piecewise continuous function on the same domain can be exactly expressed as a linear combination of sequence elements. It is a standard result in Sturm-Liouville theory and functional analysis that self-adjoint differential operators have complete eigenfunctions.}.
Some references appear at tension regarding whether the radial field equation is self-adjoint, and therefore whether or not \qnm{s} radial functions may be complete.
This apparent tension arises from the fact it is insufficient to a label a differential equation as non-self-adjoint, without additional statements about the coordinate domain, boundary conditions, and related bilinear form (e.g. inner or scalar product). 
For example, Refs.~\cite{Jaramillo:2020tuu, Leung:PhysRevA.49.3057,Nollert:1999ji} describe analogs of the radial field equation as non-self-adjoint; however, akin to Refs.~\cite{Wald:10.1063/1.524181,Beyer:10.1063/1.530922,Green:2022htq}, a key result of \PaperOne{} was the construction of a radial scalar product \textit{such that} the equation is self-adjoint.
\par Here, this scalar product is used to show that each \qnm{'s} radial function may be a member of an \textit{orthonormal basis}.
This new result places our understanding of the \qnm{}'s radial functions on the same footing as their angular functions~\cite{Cook:2014cta,OSullivan:2014ywd,London:2020uva,Berti:2014fga}.
In particular, it was shown in Ref.~\cite{London:2020uva} that each \qnm{}'s angular functions is a member of a basis, and that the properties of all such angular bases lead to the conclusion that the angular functions, namely the spheroidal harmonics with different \qnm{} frequencies, are isomorphically equivalent to the spherical harmonics, and therefore complete.
The present work takes a first step towards an analogous conclusion for the \qnm{}'s radial functions: it is shown that change-of-basis matrices between Kerr and Schwarzschild are approximately diagonal, implying that they may be isomorphically equivalent.
This communicates that, for a large range of \bh{} spins, the radial functions of Kerr may be efficiently represented using those of Schwarzschild.
\par This result and its consequences may be particularly relevant for the formulation of a time-dependent perturbation theory for \bh{}s, which presently requires the rather strong assumption that \qnm{s} are spatially complete~\cite{Green:2022htq,Cannizzaro:2023jle,Redondo-Yuste:2023ipg,Sberna:2021eui}.
\section{Scope and organization}
\bpar{Scope} The focus of this article will be on linear order curvature perturbations of isolated Kerr \bh{s}, as framed by Teukolsky, Press and Leaver~\cite{Teukolsky:1973ha,Press:1973zz,TeuPre74_3,leaver85,Leaver:1986JMP}.
\greenoff{The related version of \ee{}, namely \tk{'s} equation, is well known to be non-Hermitian~\cite{QNMTopicalReview09,Leung:PhysRevA.49.3057}, and in that context, this work comments on select aspects of problem that are shared with Hermitian problems: symmetric matrix representations, orthogonality, and the potential completeness of eigenspaces~\cite{London:2020uva,London:2021P2,London:202XP1,Green:2022htq}.}
While this article discusses only gravitational perturbations (i.e. those with spin-weight $s=\pm 2$), all calculations present make no explicit assumption about spin-weight, and in most cases have been verified to work with other physically relevant spin weights, namely $|s| \leq 2$. 
The authors do not aspire to the language of formal mathematical rigor; in place of theorems and lemmas, arguments are presented as needed, and the most technical of these are left to references.
Numerical results will use \qnm{} frequencies due to their physical relevance; however, this article's results use a framework that applies to any \textit{single frequency}.
The concurrent treatment of all \qnm{} frequencies is left to future work. 
The unit convention, $G=c=1$, will be used throughout. 
\bpar{Organization} 
This article is organized as follows.
In \sec{prelims}, a schematic background of single \bh{} perturbations is provided, culminating in the definition of the \qnm{}'s radial eigenvalue problem, and a brief review of \PaperOne{}.
In \sec{s3}, the \qnm{}'s radial eigenvalue problem is treated with \cHp{s}.
This section concludes with a discussion of the \cHp{s} pros and cons.
In \sec{s4}, this article's central result, namely the \textit{\ccHp{s}}, are introduced. 
In \sec{s5}, the \ccHp{s} are shown to tridiagonalize \tk{'s} radial equation.
In \sec{s6}, numerical results are presented to demonstrate the \ccHp{s'} end-to-end use.
\redoff{There, we note that, due to the known asymptotic properties of the QNMs radial functions' series solution, most truncated matrix representations of the radial problem will result in \textit{spectral pollution}~\cite{Nollert:1999ji,Petropoulou:2014eug,BOULTON20161,Lewin2008arXiv0812.2153L,Davies2003math2145D,MartnezAdame2007}.}
In \sec{discus}, the results of the preceding sections are discussed in the context of past and future work. 
%
%
\section{Preliminaries}
\label{prelims}
\bpar{Observables and perturbative quantities} A \gw{} observatory will  detect a linear combination of the two strain polarizations present in Einstein's \gr{}, $h_+$ and $h_\times$~\cite{Blanchet:2013haa}.
We may parameterize each using Boyer-Lindquist coordinates, $\{t,r,\theta,\phi\}$ where, asymptotically far from a radiating source, $t$ may be defined to measure time elapsed for an observer, $r$ the luminosity distance, and $(\theta,\phi)$ the usual spherical polar angles~\cite{BoyerLindquist:1967}. 
Since all $\theta$ dependence is in the form of $u=\cos(\theta)$, $u$ will henceforth be used~\cite{leaver85}. 
\par From the perspective of \te{}, it is convenient to work with the curvature scalar $\psi_4$ which is related to the strain polarizations via
\begin{subequations} 
    \label{p1}
    \begin{align}
        \label{p1a}
        h \; &= \; h_+ \, - \, i \, h_\times \;
        \\
        \label{p1b}
             &= \; \int_{-\infty}^{t} \int_{-\infty}^{t'} \psi_4(t'',r,u,\phi) \; dt''   dt'  \; .
    \end{align}
\end{subequations} 
In \eqn{p1b}, $\psi_4$ is the fifth scalar of the Weyl tensor in the Newman-Penrose formalism~\cite{NP62}.
It is a field of spin-weight $s=-2$, and encodes all relevant information about gravitational radiation~\cite{NP62,NP66,Ruiz:2007yx,Merlin:2016boc}.
For a single isolated \bh{}, the rescaled Weyl scalar
\begin{align}
    \label{p2}
    \psi \; &= \; -( r - i a u )^{-4} \; \psi_4 \;.
\end{align}
satisfies \te{}~\cite{Teukolsky:1973ha,Press:1973zz,TeuPre74_3}. \redoff{In \eqn{p2}, $a$ is the \bh{} spin magnitude divided by the \bh{} mass.}
Schematically, if we denote the second order linear partial differential operator within \te{} as $\LMasterB$, then for isolated black holes,
\begin{align}
    \label{p3}
    \LMasterB \; \psi \; = \; 0 \;.
\end{align}
Since $\LMasterB$ has partial derivatives on each spacetime coordinate, the solutions $\psi$ are waves~\cite{Teukolsky:1973ha}.
\bpar{Quasinormal Modes} The \qnm{s} of an isolated Kerr \bh{} comprise a sequence of solutions to \eqn{p3} that are ingoing waves at the \bh{} horizon and purely outgoing radiation towards spatial infinity~\cite{Teukolsky:1973ha,leaver85}.
These conditions on a \qnm{'s} phase velocity are known to constrain the functional form of $\psi$ by defining the allowed radial behavior asymptotically near the event horizon and spatial infinity~\cite{Cook:2014cta,Press:1973zz,leaver85}.
\par An individual \qnm{} has spatiotemporal dependence given by the product of four functions,
\begin{align}
    \label{p4}
    \psi \; \propto \; R(r) \, S(u) \, e^{-i \cw t} \, e^{-i m \phi}\;.
\end{align}
In \eqn{p4}, $R(r)$ will be referred to as simply a ``radial function'' and $S(u)$ a spheroidal harmonic~\cite{Teukolsky:1973ha}.
Both are \textit{implicitly parameterized} by the field's spin weight, $s$, as well as the azimuthal and temporal mode numbers, $m$ and $\cw$ respectively.
For given values of $\cw$ and $m$, the study of possible solutions for each $R(r)$ and $S(u)$ result in two additional mode numbers, $n$ and $\ell$, where $\ell$ is known to be closely related to the order of Jacobi polynomials~\cite{London:202XP1,Fackerell:1977,London:2020uva}. 
One goal of the present work is to provide an analogous mathematical understanding of the colloquially named ``overtone'' number, $n$.
\par In \eqn{p4}, $\cw \; = \; \omega \, - \, i/\tau$ is a complex valued \qnm{} or ``ringdown'' frequency, and $\tau$ must be positive for time domain stability of the \bh{}~\cite{Press:1973zz,leaver85,Whiting:1989ms}. 
Consequently, in the time domain, each \qnm{} is a damped sinusoid with central frequency $\omega$, and damping time $\tau$.
\te{} constrains the functional form of each \qnm{}, and the details of the perturber constrain its overall excitation amplitude~\cite{Leaver86c,Berti:2006:ExFacs}.
\par Furthermore, application of \eqn{p4} to \te{} yields that \eqn{p3} separates (with separation constant $A$) into  
\begin{subequations} 
    \label{p5}
    \begin{align}
        \label{p5a}
        \mcL{_u} \, S(u) \; &= \; - A \, S(u) \; ,
        \\
        \label{p5b}
        \mcL{_r} \, R(r) \; &= \; + A \, R(r) \; .
    \end{align}
\end{subequations} 
Henceforth, \eqn{p5a} and \eqn{p5b} will be referred to as \tk{}'s angular and radial equation, respectively.
They are eigenvalue equations, where $-A$ and $+A$ are the respective complex valued\footnote{The complex nature of the eigenvalues is a consequence of the fact that \eqnsa{p5a}{p5b} have complex valued terms. It will be seen in \eqn{p14} that this coincides with a non-conjugate symmetric scalar product.} eigenvalues~\cite{London:202XP1,London:2020uva,Fiziev:2009,Leaver:1986JMP}.
Therefore, $S(u)$ and $R(r)$ are the respective eigenfunctions.
In \eqn{p5}, $\mcL_u$ and $\mcL_r$ are second order linear differential operators that will be referred to as \tk{}'s angular and radial operators, respectively. 
Each implicitly depends on $s$, $m$ and $\cw$, as well as the \bh{} mass $M$, and spin $a=|\vec{J}|/M$.
\par Considered on their own, i.e. outside of the explicit physical context described thus far where $\cw$ is assumed to be a \qnm{} frequency, \eqnsa{p5a}{p5b} are known to have a discrete space of eigenfunctions, which may be labeled in ${\ell'}$ and ${n'}$, respectively~\cite{leaver85,Berti:2007zu}. 
Their eigenvalues implicitly inherit these labels, 
\begin{figure}[t]
    
    
    \begin{tabular}{c}
        \hspace{-0.6cm}\includegraphics[width=0.42\textwidth]{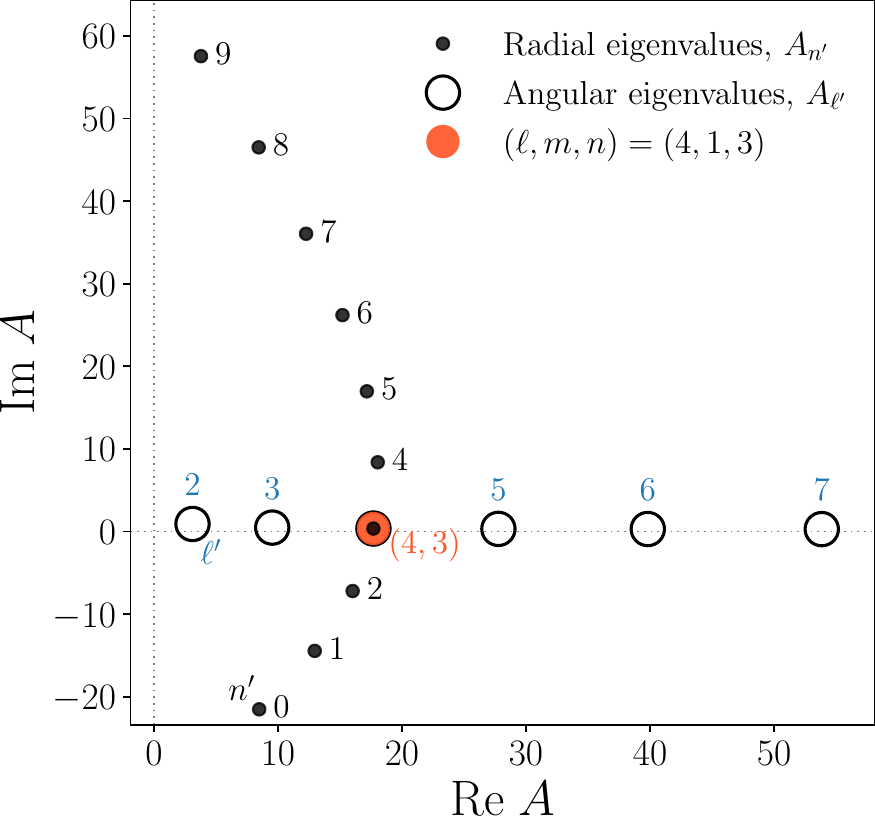}
    \end{tabular}
    
    \caption{ Distributions of radial (black dots) and angular eigenvalues (open circles) for $s=-2$, \bh{} dimensionless spin $a/M=0.7$, and the $(\ell,m,n)=(4,1,3)$ \qnm{} with $M\cw_{413}=0.8545-0.6305i$. 
    Since both eigenvalue distributions have been computed with a \qnm{} frequency, they have one coincident value corresponding to $(\ell',n')=(\ell,n)=(4,3)$. 
    Radial egenvalues shown have been computed using the results of the current work, and verified using Leaver's method~\cite{leaver85}.
    }
    \label{F1} 

\end{figure}
\begin{subequations} 
    \label{p6}
    \begin{align}
        \label{p6a}
        \mcL'\hspace{-2.5pt}{_u} \, S_{{\ell'}}(u) \; &= \; - A_{{\ell'}} \, S_{{\ell'}}(u) \; ,
        \\
        \label{p6b}
        \mcL'\hspace{-2.5pt}{_r} \, R_{{n'}}(r) \; &= \; + A_{{n'}} \, R_{{n'}}(r) \; .
    \end{align}
\end{subequations} 
\par \Eqnsa{p6a}{p6b} hold for \textit{any} frequency parameter $\cw'$ that is independent of $\ell'$ and $n'$.
Since $\mcL{_u}$ and $\mcL{_r}$ depend on the frequency parameter, they are also primed in \eqn{p6}.
It is \textit{only} for the \qnm{} frequencies that $A_{\ell'}=A_{n'}=A$, and \eqn{p5} is recovered.
Since $\ell'$ and $n'$ are independent, for any one \qnm{} frequency, there will be only one value of $\ell'$ and one value of $n'$ such that $A_{\ell'}=A_{n'}$.
These special labels are those for which $\ell'=\ell$ and $n'=n$.
For these reasons, the \qnm{} frequencies and their separation constants are often labeled with spatial mode numbers, i.e. $\cw_\lmn{}$ and $A_\lmn$ respectively~\cite{Berti:2005ys,London:2014cma,Berti:2016lat,MaganaZertuche:2021syq}.
As a result, the \qnms{} are a non-trivial infinite sequence, wherein each element satisfies \eqn{p6} with a nominally distinct frequency parameter, $\cw'=\cw_\lmn$~\cite{London:2020uva}.  
\par An example of the angular and radial eigenvalue distributions is shown in \Fig{F1}.
There, the typical qualitative features of each distribution are shown: the angular eigenvalues, $A_{\ell'}$, will typically be spaced horizontally along the real axis, and the radial eigenvalues, $A_{n'}$, will vary most strongly in their imaginary part.
Since the case shown has been constructed using a \qnm{} frequency, there will be a single coincident eigenvalue, which is the one of physical interest~(i.e. the one compatible with the construction of \ceqn{p5}).
\bpar{Notation} For convenience of notation, \qnm{} frequencies will henceforth be referred to as $\cw_\lmn$, and $\cw$ will refer to an arbitrary frequency parameter.
The related radial eigenvalue relation will be referred to simply as
\begin{align}
    \label{p7}
    \mcL{_r} \, R(r) \; &= \; A \, R(r) \; ,
\end{align}
where it is implicitly understood that the eigenvalues, $A$, and eigenfunctions, $R(r)$, are elements of that equation's discrete spectra.
A transformed version of the radial equation will be encountered shortly. 
At that time, the same notational strategy will be used. 
\bpar{A radial scalar product} The angular equation, \eqn{p6a}, is well-known to be of the Sturm-Liouville type (i.e. $\mcL_u$ is self-adjoint)~\cite{London:2020uva,London:202XP1,Courant1954,lax2002functional}.
Consequently, its eigenfunctions, $S_{\ell'}(u)$, are complete\footnote{\greenoff{See Ref.~\cite{Finster:2015xma} for a discussion of completeness in the context of single complex valued oblateness parameters (i.e. $a \cw $), and see Ref.~\cite{London:2020uva} for related discussion of completeness in the \qnm{} context, where oblateness parameters differ from mode to mode.} } and orthogonal with respect to a scalar product, $\brak{a}{b}$, defined by the integral of $a(u)b(u)$ on $u\in [-1,1]$. 
The application of this angular scalar product is fundamental to the use of spherical and spheroidal harmonics in \gw{} theory and data analysis~\cite{Blanchet:2013haa,Khan:2015jqa,Garcia-Quiros:2020qpx}.
\par In contrast, the radial equation, \eqn{p6b}, is known to not explicitly be of the Sturm-Liouville type.
This simply means that, in practice, the application of Sturm-Liouville theory must begin with the construction of a \textit{weighted} scalar product, under which $\mcL_r$ is self-adjoint~\cite{Courant1954,pinchover_rubinstein_2005,Kristensson:2010}. 
Below, we provide a brief overview of the construction and evaluation of such a radial scalar product. We refer the reader to \PaperOne{} for a complete pedagogical development.
\par Using the compactified radial coordinate, 
\begin{align}
    \label{p8}
	\xi \; &= \; (r-\rp)/(r-\rm) \; ,
\end{align}
and applying the \qnm{s'} asymptotic boundary conditions via an appropriate rescaling, 
\begin{align}
    \label{p9}
    R(r(\xi)) \; = \; \mu(\xi) f(\xi)\;,
\end{align}
the radial problem is transformed so that the radial function of interest is now $f(\xi)$, and differential operator $\mcL_r$ becomes 
\begin{subequations}
    \label{p10}
    \begin{align}
        \label{p10a}
        \mcL{_\xi}  \; &= \; (\text{C}_0+\text{C}_1 (1-\xi ))
        \\
        \label{p10b}
        & + \; \left(\text{C}_2+\text{C}_3 (1-\xi )+\text{C}_4 (1-\xi )^2\right)\partial_{\xi} 
        \\
        \label{p10c}
        & + \; \xi  (\xi -1)^2 \partial_{\xi}^2 \; .
    \end{align}
\end{subequations}
The operator $\mcL_\xi$ may be compactly written by defining the operator $\mcD{}_\xi$ to be the terms within \eqnsa{p10b}{p10c}, 
\begin{align}
    \label{p11}
    \mcL_\xi \; = \; (\text{C}_0+\text{C}_1 (1-\xi )) \, + \, \mcD_\xi \; .
\end{align}
\par Note that in \eqn{p8}, $\xi$ maps $r\in[r_+,\infty)$ to $\xi\in[0,1)$, thereby compactifying the radial coordinate.
In \eqn{p9}, $\mu(\xi)$ is comprised of factors related to the singular exponents of $\mcL_r$.
Its explicit form is given in Eq.~(12) of \PaperOne{}.
In \eqn{p10}, $\tx{C}_0$ through $\tx{C}_4$ are complex valued constants determined by the \bh{} mass $M$, the \bh{} spin $a$, the mode numbers $\{\cw,m,l,n\}$, and the field's spin weight $s$:
\begin{subequations}
    \label{p12}
    \begin{align}
        \label{p12b}
        \text{C}_0 \; &= \; -2 a m \cw-2 i \cw (-\delta +M(2  s+1))
        \\ \nonumber
        & \quad \;\; +\cw^2 (\delta +M) (\delta +7 M) \; ,
        \\
        \label{p12c}
        \text{C}_1 \; &= \; 8 M^2 \cw^2+ s (4 i M \cw-1)+6 i M \cw-1
        \\ \nonumber 
        & \quad \;\; - (4 M \cw+i) \frac{\left(a m-2 M^2 \cw\right)}{\delta } \; ,
    \end{align}
    \begin{align}
        \label{p12d}
        \text{C}_2 \; &= \; 4 i \delta  \cw \; ,
        \\
        \label{p12e}
        \text{C}_3 \; &= \; -2(s+1)+ 4 i \cw (M-\delta ) \; ,
        \\
        \label{p12f}
        \text{C}_4\; &= \; s+3-6 i M \cw + \frac{i \left(a m-2 M^2 \cw\right)}{\delta } \; ,
    \end{align}
    \begin{align}
        \label{p12a}
        \delta \; &= \; \sqrt{ M^2 - a^2 } \; .
    \end{align}
\end{subequations}
Lastly, note that none of $\tx{C}_0$ through $\tx{C}_4$ become zero when $a=0$.
As a result, \eqn{p11} is of the confluent Heun type for both Schwarzschild and Kerr~\cite{London:202XP1}.
\par Together, \eqns{p8}{p12} allow the radial eigenvalue relation~(\ceqn{p7}) to be rewritten as
\begin{subequations}
    \label{p13}
    \begin{align}
        \label{p13a}
        \mcL{_\xi} \, f(\xi) \; &= \; A \, f(\xi) \; .
    \end{align}
The equivalent statement in standard bra-ket notation is
    \begin{align}
        \label{p13b}
        \mcL{_\xi} \, \ket{f} \; &= \; A \, \ket{f} \; .
    \end{align}
\end{subequations}
In \eqn{p13b}, and henceforth, the use of bra-ket notation is consistent with the fact that $f(\xi)$ is a vector in space whose bilinear product is the radial scalar product of interest. 
The practical use of kets, e.g. $\ket{\Phi}$, and bras, $\bra{\Phi}$, will be demonstrated in the context of the scalar product. 
\par A scalar product between two smooth functions of $\xi$, $a(\xi)$ and $b(\xi)$, is defined as
\begin{align}
    \label{p14}
    \brak{\mathrm{a}}{\mathrm{b}} \; = \; \int_{0}^{1} \, \mathrm{a}(\xi) \, \mathrm{b}(\xi) \; \tx{W}(\xi) \; d\xi \; ,
\end{align}
where $\tx{W}(\xi)$ is the product's \textit{weight function}.
In turn, $\tx{W}(\xi)$ is defined such that the operator, $\mcL_\xi$, is equal to its adjoint, $\adj{\mcL}_\xi$, with respect to the scalar product: 
\begin{align}
    \label{p15}
    \brak{a}{\mcL_\xi b} \; &= \; \brak{\adj{\mcL}_\xi a}{b} \; 
    \\ \nonumber
    \; &= \; \brak{{\mcL}_\xi a}{b} \; .
\end{align}
Note that there is no conjugation in \eqn{p14}. As a result, \eqn{p15}'s adjoint operator, $\adj{\mcL}_\xi$, does not involve complex conjugation.
\par By combining the definition of $\mcL_\xi$~(\ceqn{p10}) with the requirement of self-adjointness~(\ceqn{p15}) and then applying integration by parts, it may be shown that
\begin{align}
    \label{p16}
       \tx{W}(\xi) \; &= \;  \xi ^{\text{B}_0} (1-\xi )^{\text{B}_1} e^{\frac{\text{B}_2}{1-\xi }} \;,
\end{align}
where constants $\tx{B}_0$ through $\tx{B}_2$ are,
\begin{subequations}
    \label{p17}
    \begin{align}
        \label{p17a}
        \text{B}_0 \; &= \; \text{C}_2+\text{C}_3+\text{C}_4-1
        \\
        \label{p17b}
        \text{B}_1 \; &= \; -\text{C}_2-\text{C}_3-2
        \\
        \label{p17c}
        \text{B}_2 \; &= \; \text{C}_2 \; .
    \end{align}
\end{subequations}
\par In practice, if the series expansion of two functions, $\tx{a}(\xi)$ and $\tx{b}(\xi)$, exist and are known,
\begin{align}
    \label{p18}
    \tx{a}(\xi) \; = \; \sum_{j=0} \, \text{a}_j \, \xi^j \quad\text{  and  }\quad  \tx{b}(\xi) \; = \; \sum_{k=0} \, \text{b}_k \, \xi^k \;  ,
\end{align}
then the scalar product may be evaluated as 
\begin{align}
    \label{p19}
    \brak{\tx{a}}{\tx{b}} \; = \; \sum_{j,k} \; \text{a}_j \, \text{b}_k \, \left< \xi^{j+k} \right> \; .
\end{align}
In \eqn{p19}, $\left< \xi^{j+k} \right>$ are called \textit{monomial moments}~\cite{chihara2011introduction}. 
Using the definition of the weight function, it has been observed (e.g.~\cite{Cho:2009wf,Leaver86c,abramowitz+stegun}) that each monomial moment $\left< \xi^p \right>$ can be evaluated using the Gamma function, $\Gamma(x)$, and the Tricomi confluent hypergeometric function, $U(x,y,z)$,
\begin{align}
    \label{p20}
    \nonumber
    \left< \xi^{p} \right> \; = \; e^{\text{B}_{2}} \, \Gamma (&\text{B}_{0}+p+1) 
    \\
    &\times \; U(1+\text{B}_{0}+p,-\text{B}_{1},-\text{B}_{2}) \; . 
\end{align}
\Eqnsa{p19}{p20} enable the scalar product to be evaluated for complex values of $\tx{B}_0$, $\tx{B}_1$ and $\tx{B}_2$ via the standard analytic continuations of $\Gamma(x)$ and $U(x,y,z)$~\cite{London:202XP1}.
\bpar{Confluent Heun polynomials} In many areas of physics, polynomial solutions to a system's equations of motion play an important role in describing that system's natural modes~\cite{Sakurai1993Modern,Cook:2014cta,London:2020uva,Fackerell:1977}.
For the \qnm{'s} radial problem~(\ceqn{p13}), one approach to understanding polynomial solutions begins by studying the differential part of $\mcL_\xi$, namely
\begin{align}
    \label{p21}
    \mcD{_\xi}  \; &= \; \left(\text{C}_2+\text{C}_3 (1-\xi )+\text{C}_4 (1-\xi )^2\right)\partial_{\xi}
    \\ \nonumber 
    & \hspace{.5cm} + \; \xi  (\xi -1)^2 \partial_{\xi}^2  \; .
\end{align}
This operator defines polynomials that satisfy a two-parameter eigenvalue problem,
\begin{subequations}
    \label{p22}
    \begin{align}
        \label{p22a}
        \mcD{_\xi} \; \ket{y_{pk}}  \; &= \; (\lambda_{pk} \; + \; \mu_{p} \, \xi) \; \ket{y_{pk}} \;,
    \end{align}
    where 
    \begin{align}
        \label{p22b}
        \mu_{p} \; &= \; p\,(\,p+\text{C}_4-1\,) \; .
    \end{align}
\end{subequations} 
In \eqn{p22a}, $\lambda_{pk}$ is the eigenvalue, and the linear factor, $\mu_p$, is required for the polynomials, $\ket{y_{pk}}$, to terminate after $p+1$ terms~\cite{London:202XP1,ronveaux1995heun}.
For each polynomial order $p$, there are $p+1$ solutions for $\lambda_{pk}$ that allow for termination~\cite{NIST:DLMF:ConfHeunPoly,London:202XP1}. 
The label $k$ heuristically orders these solutions according to the imaginary part of, $\tx{Im}\,\lambda_{pk}$~\cite{London:202XP1}.
Since \eqn{p22a} is a confluent Heun equation, $\ket{y_{pk}}$ comprise a class, or multiplex, of order-$p$ \cHp{s}~\cite{ronveaux1995heun}.
\par The ostensibly unusual (two parameter) nature of \eqn{p22a} is a consequence of $\mcD_\xi$'s polynomial coefficient functions~(\ceqn{p21}).
For example, when acting upon any polynomial of order $p$, $\xi(1-\xi)^2\partial_{\xi}^2$ will generally result in a polynomial of order $p+1$.
The same is true of $\left(\text{C}_2+\text{C}_3 (1-\xi )+\text{C}_4 (1-\xi )^2\right)\partial_{\xi}$.
Therefore, $\mcD_\xi$ generally increases the order of polynomials by one, which precludes the usual eigenrelation, where order is preserved. 
\par A consequence of the multi-parameter eigenrelation is that the \cHp{s} do not obey exactly the same kind of orthogonality found in the classical polynomials.
This may be seen by noting that, like $\mcL_\xi$, $\mcD_\xi$ is self-adjoint \wrt the scalar product,
\begin{align}
    \label{p23}
    \brak{y_{pk'}}{\mcD_\xi y_{pk}} \; = \; \brak{\mcD_\xi y_{pk'}}{y_{pk}} \; .
\end{align}
Applying the eigenvalue relationship, \eqn{p22a}, to both sides of \eqn{p23} yields 
\begin{align}
    \label{p24}
    \nonumber
    (\lambda_{pk}-\lambda_{p'k'})\,&\brak{y_{p'k'}}{y_{pk}}& 
    \\ 
    \; &= \; (\mu_{p'}-\mu_{p}) \,\bra{y_{p'k'}} \xi \; \ket{y_{pk}} \; .
\end{align}
For self-adjoint and single-parameter eigenvalue problems, the \rhs{} of \eqn{p24} is identically zero, resulting in the well-known orthogonality of eigenfunctions with unique eigenvalues~\cite{Axler:2015,Sakurai1993Modern}.
Since $\mu_p$ is not generally zero, \textit{only} the polynomials of like order (i.e. $\mu_{p'}-\mu_p=0$) are orthogonal,
\begin{align}
    \label{p25}
    \brak{y_{pk'}}{y_{pk}}  \; &= \; \delta_{k'k} \; .
\end{align}
In \eqn{p25}, $\ket{y_{pk}}$ are normalized, i.e. scaled such that $\brak{y_{pk}}{y_{pk}}=1$~\cite{London:202XP1}.
\par It was observed in \PaperOne{} that $\lambda_{pk}$ are non-degenerate across physically relevant values of \bh{} mass and spin.
Consequently, each $p+1$ dimensional vector space of polynomials $y_{pk}$ is expected to be a basis, meaning that the identity operator for order-$p$ polynomials may be written as,
\begin{align}
    \label{cHpI}
    \I_{p} \; = \; \sum_{k=0}^p \, \ketbra{y_{pk}}{y_{pk}} \; .
\end{align}
However, since each basis for order $p$ polynomials spans all previous bases, the total set of confluent Heun polynomials satisfying \eqn{p22a} is redundant, or rather \textit{overcomplete}~\cite{Christensen2003}.
\redoff{
    \par Lastly, the overcompleteness of the confluent Heun polynomails does not imply that it is trivial to use them as a basis for general solutions to the \qnm{} radial problem. Doing so presents two challenges that we will discuss in upcoming sections. The first challenge is that each fixed-order confluent Heun polynomial basis need not have well ordered expansion coefficients, meaning that there is \textit{not necessarily} an order to how much each polynomial contributes to a radial expansion. This challenge will be discussed in the upcoming section, \csec{s3}.
    \par The second challenge is that the matrix representation of e.g. $\mcL_\xi$ in a finite dimensional basis need not have \textit{exactly the same spectra} as its infinite dimensional counterpart (i.e. there may appear modes that are purely the effect of truncation)~\cite{Petropoulou:2014eug}. This challenge will be revisited in \sec{s6}.
}
\section{The radial problem with Confluent Heun Polynomials}
\label{s3}
\smallskip
\par The properties of \cHp{s} have three implications for each \qnm{'s} radial problem~(\ceqn{p13b}).
\par The first is that the radial problem, when represented in a basis of order-$p$ \cHp{s}, may written simply, with a \textit{deformation parameter}, 
\begin{align}
    \label{alpha-p}
    \alpha_{p} \; &= \; \mu_{p}-\tx{C}_1 \; ,
\end{align}
that determines the size of all off-diagonal terms. 
Concurrently, off diagonal terms are proportional to $\brak{y_{pk'}}{\xi\,|y_{pk}}$, meaning that, for non-zero $\alpha_p$, the matrix representation of the radial problem may not be approximately diagonal in the sense that elements do not necessarily decrease far away from $k'=k$.
\par The second implication is that, due to properties of $\alpha_p$, a kind of \textit{duality} exists in the radial problem's solution space.
Most simply, if $\alpha_p=0$ and $p$ is a \nnint{}, then $p+1$ solutions will be polynomials, while the remaining infinity of solutions will be infinite series.
\par The final implication is that the physical \qnm{s} will not generally be approximately polynomial in nature.
Along with the aforementioned duality, this implies that there will typically be two, sometimes overlapping, sectors of solution space.
It happens that this concurrently highlights practical issues that arise when attempting to approximate a function with an arbitrary fixed-order polynomial basis.   
\par While these implications provide an instructive reference for the basic properties of non-polynomial solutions to the radial equation, they also motivate the development of a basis comprised of uniquely ordered polynomials. 
That is the topic of the next section.  
For now, the radial problem with \cHp{s} is to be discussed in more detail. 
\smallskip
\bpar{The radial problem} Any order-$p$ \cHp{} basis may be used to approximate the radial problem,
\begin{subequations}
    \label{cHpApx}
    \begin{align}
        \label{cHpApx-a}
        \I_{p} \, \mcL{_\xi} \,\I_{p}\, \ket{f} \; &= \; A \, \I_{p} \, \ket{f} \; ,
        \\
        \label{cHpApx-b}
        \LcH \, \vec{\tx{f}} \; &= \; A \, \vec{\tx{f}} \; .
    \end{align}
\end{subequations}
In \eqn{cHpApx-a}, insertion of the identity operator~(\ceqn{cHpI}) signals that e.g. $\ket{f}$ will be represented in the basis of order-$p$ \cHp{s}.
In \eqn{cHpApx-b}, \eqn{cHpI} has been applied, and it has been recognized that the equation is equivalent to a matrix eigenvalue equation, where $\LcH$ is a $(p+1)\times (p+1)$ matrix with elements, 
\begin{align}
    L^{(\tx{cH})}_{p,\,k'k} \; = \; \brak{y_{pk'}}{\mcL_\xi \,|\,y_{pk}} \; ,
\end{align}
and $\vec{\tx{f}}$ has elements $\brak{y_{pk}}{f}$.
\par In many cases, $\LcH$ will have many off-diagonal terms.
This may be understood by recalling that $\mcL_{\xi}=(\tx{C}_0+\tx{C}_1 [1-\xi])+\mcD_\xi$~(i.e. \ceqn{p11}), and then applying the \cHp{} eigenvalue relationship to $\brak{y_{pk'}}{\mcL_\xi \,|\,y_{pk}}$, yielding
\begin{subequations}
    \label{LL}
    \begin{align}
        \nonumber
        \brak{y_{pk'}}{\mcL_\xi \,|\,y_{pk}} \; = \; &(\tx{C}_0+\tx{C}_1+\lambda_{pk}) \, \delta_{k'k} \, 
        \\ \label{LL-a}
        &+ \,\alpha_p\, \brak{y_{pk'}}{\xi\,|\,y_{pk}} \; ,
    \end{align}
\end{subequations}
In \eqn{LL-a}, $\alpha_p$ is defined by \eqn{alpha-p}.
\par If $\alpha_p=0$, \eqnsa{cHpApx-b}{LL-a} yield that the \qnm{} separation constant is simply given by 
\begin{align}
    \label{cHpA}
    A_{pk} \; = \; \CZero+\COne+\lambda_{pk}\; .
\end{align} 
If a \qnm{'s} radial functions is well approximated by a single \cHp{}, then \eqn{cHpA} will be similarly approximate. 
\par When $\alpha_{p} \neq 0$, the off-diagonal elements of $\LcH$ will be proportional to $\brak{y_{pk'}}{\xi \,|\,y_{pk}}$.
It was found in \PaperOne{} that $|\brak{y_{pk'}}{\xi \,|\,y_{pk}}|$ do not typically decrease in value as $|k'-k|$ increases.
\redoff{Consequently, $\LcH$ will typically be neither band-diagonal nor approximately diagonal when $\alpha_p\neq0$.}
\bpar{Polynomial / non-polynomial duality}
From inspection of \eqn{LL-a}, it is clear that when $\alpha_p=0$, $\LcH$ is diagonal.
This implies that $\LcH$ may be \textit{approximately diagonal} when $|\alpha_{p}|\ll 1$. 
\redoff{In the case where $\alpha_p=0$, a finite subset of the solution space will be comprosed of polynomails, and the remainder will be an infinite set of non-polynomial functions.
\par It happens that, when $\alpha_p \neq 0$, the solution space will typically be comprised of two qualitatively destict sectors: some solutions will be approximately polynomial in the sense that their series expansions converge quickly after some particular term, and other strongly non-polynomialsolutions whose series expansions converge relatively slowly (see e.g. the asymptotic analysis in Ref.~\cite{Nollert:1999ji}). 
Henceforth, we refer to the presence of these two solutions types as ``{polynomial / non-polynomial duality}\footnote{\redoff{We expect ``polynomial / non-polynomial duality'' to be related to the existence of quasi-bound states for $s=0$~\cite{Brito:2015oca,Vieira:2021nha}.}}''.}
\par This concept may be refined by defining \textit{pseudo-polynomial orders}, $\psp$ and $\psm$, such that $\alpha_{\pspm}$ is zero: 
\begin{subequations}
    \label{pspConstraint}
    \begin{align}
        \label{pspConstraint-a}
       \alpha_{\pspm} \; &= \; \mu_{\pspm}- \tx{C}_1
       \\ 
       \label{pspConstraint-b}
       \; &=  \,\pspm\,(\,\pspm+\tx{C}_4\,-\,1\,)\,- \;\tx{C}_1 =  \; 0 \; .
    \end{align}
\end{subequations}
In \eqn{pspConstraint-b}, $\mu_{\pspm}$ has been rewritten using \eqn{p22b}.
Solving for $\pspm$, one finds that 
\begin{align}
    \label{pstar}
    \pspm = \frac{1}{2} \left[ (1-\tx{C}_4) \pm \sqrt{4 \text{C}_1+(1-\tx{C}_4)^2} \right] \;.
\end{align}
%
%
In what follows, $\psp$ wll be referred to as the \textit{upper} pseudo-polynomial order, and $\psm$ will be referred to as the \textit{lower} pseudo-polynomial order.
\def\refpsp{6.2586+0.0743i}
\begin{table}[t]
    \begin{tabular}{|>{\centering\arraybackslash}p{1cm}|>{\centering\arraybackslash}p{3.5cm}|>{\centering\arraybackslash}p{3.5cm}|}
        \hline
        {$n$} & {$\psp$}  & {{$\psm$}}
        \\ 
        \hline
        \hline
        {$0$} & $1.3879+0.5964i$ & $\!\!\!\!-0.6768+2.1304i$ 
        \\
        \hline
        {$1$} & $2.1725+0.5415i$ & $\!\!\!\!-0.0230+2.0846i$
        \\
        \hline
        {$2$} & $2.9791+0.4394i$ & $0.6490+1.9996i$ 
        \\
        \hline
        {$4$} & $4.6187+0.1537i$ & $2.0152+1.7615i$ 
        \\
        \hline
        {$6$} & $\refpsp$ & $3.3816+1.6954i$ 
        \\
        \hline
        {$8$} & {---} & {---} 
        \\
        \hline
        {$10$} & {$11.0503+0.0551i$} & {$7.3742+1.6794i$} 
        \\
        \hline
        {$12$} & {$13.0252+0.0344i$ } & {$9.0198+1.6621i$} 
        \\
        \hline
        {$14$} & {$15.0134+0.0152i$}  & {$10.6765+1.6461i$} 
        \\
        \hline
        {$16$} & {$17.0123-0.0004i$}  & {$12.3420+1.6331i$} 
        \\
        \hline
        {$18$} & {$19.0204-0.0084i$} & {$14.0152+1.6265i$} 
        \\
        \hline
        {$20$} & {$21.0298-0.0007i$} & {$15.6895+1.6329i$} 
        \\
        \hline
    \end{tabular}
    \caption{ Select pseudo-polynomial orders, $\psp$ and $\psm$, for \bh{} spin of $a/M=0.7$, and \qnm{} indices $(s,\ell,m)=(-2,2,2)$. Parameters for select \qnm{} overtone numbers, $n$, are shown. \greenoff{Only even $n$ are shown for brevity, and results for odd $n$ show the same qualitative trend (e.g. for $n=13$, $(\psp,\psm)=(14.0178+0.0245i,9.8469+1.6539i)$). We do not show the result for $n=8$, where further conditions are needed due to the appearance of non-traditional modes~\cite{Cook:2014cta}.}}
    \label{pspm-examples}
\end{table}
\par For the moment, we are concerned with the implications of the pseudo-polynomial orders on solutions to the radial problem (\ceqn{cHpApx}), the \qnm{s} radial functions, and the ultimate practicality of \cHp{} representations for non-polynomial problems. 
\par \Eqn{pstar} implies a number of simple relationships between $\alpha_p$, $\mu_p$, $\psp$ and $\psm$.
For example, \eqn{pstar} may used to show that 
\begin{align}
    \label{psm-2}
    \psp + \psm  \; &= \; 1-\CFour \; .
\end{align}
It also follows that $\COne=-\psp \psm$.
In terms of physical parameters~(\ceqn{p12}), $\psp$ and $\psm$ are
\begin{subequations}
    \label{pspm}
    \begin{align}
        \label{pspm-a}
        \psp \; &= \; -1 \, + \, 4i M \cw \; ,
        \\
        \label{pspm-b}
        \psm \; &= \; \psp \, -s \, - \, 2 i M \cw \, - \, \frac{i \left(a m-2 M^2 \cw\right)}{\delta } \; .
    \end{align}
\end{subequations}
\redoff{In \eqn{pspm} we have simplified \eqn{pstar} by adopting the strategy of Ref.~\cite{Teukolsky:1973ha}, and holding that if $x$ is a complex number, then e.g. $\sqrt{-x^2}=i\,x$, rather than the generally correct value of $-i\,x\,\tx{sign}{( \tx{Im}(x))}$. The effect of this choice is to swap $\psp$ with $\psm$. Thus, it should be understood that \eqnsa{pspm}{pstar} are formally different.}  
\par Lastly, \eqn{pstar} allows $\alpha_p$ to be written as
\begin{align}
    \label{alpha-p-factored}
    \alpha_{p} \; = \; (p-\psp)(p-\psm) \; .
\end{align}
\par \redoff{From \eqn{psm-2}, three scenarios may be deduced for which the solution space includes a finite number of polynomials in addition to an infinity of non-polynomials:
\begin{itemize}
    \item[\textit{(i)}] The first scenario is that, if $\psp$ is a positive integer, then $\psm$ typically will not be unless $\CFour$ is a sufficiently negative integer.
    \item[\textit{(ii)}] The second scenario is simply that of the first, but with $\psm$ being a positive integer, and $\psm$ having a fractional part.
    \item[\textit{(iii)}] In the remaining scenario \textit{both} $\psm$ and $\psp$ are integers.
    In this case, it is possible for the solution space to support two sets of polynomials, each of different order, as well as an infinite number of non-polynomial solutions.
\end{itemize}
%
%
}
%
%
\par Generally, neither $\psp$ nor $\psm$ will be positive integers for the \qnm{s}, and their proximity to a positive integer will vary with different values of the overtone index, $n$.
In particular, solutions to the \qnm{} radial problem may only be closely associated with \cHp{s} when $\cw_\lmn$ are such that $\psp$ or $\psm$ is close to a positive integer according to
\redoff{
\begin{subequations}
    \label{near-int}
    \begin{align}
        \label{near-int-a}
        {|\tx{Im}\, \pspm }/{\tx{Re}\,\pspm}| \; \ll \; 1 \; ,
    \end{align}
and
    \begin{align}
        \label{near-int-b}
        \pspm-\left[ \tx{Re}\,\pspm \right] \; \ll \; 1 \; .
    \end{align}
\end{subequations}
In \eqn{near-int}, $\left[ x \right]$ denotes rounding to the nearest integer~(e.g. $\left[ 3.1416 \right]=3$).}
\par \Tbl{pspm-examples} provides examples of $\pspm$ for select \qnm{} cases with mode numbers $(\ell,m)=(2,2)$.
\redoff{Values of $\pspm$ have been defined using \eqn{pstar}.}
It is observed that the \qnm{'s} $\psp$ only very roughly tracks the overtone index, being approximately off by one for e.g. $n=2$, but off by significantly less than one for $n=6$ \greenoff{and above}.
%
%
%
The case of the fundamental \qnm{}, $n=0$, demonstrates a general observation, namely that the $n=0$ \qnm{s} are typically not close to \cHp{} systems in that their $\pspm$ are not close to integers.
%
%
%
%
\bpar{Polynomial noise} Since the \qnm{s} $\pspm$ are typically complex valued, it is not guaranteed that using an order-$p$ \cHp{} basis with e.g. $p = \lfloor\max(|\psp|,|\psm|)\rfloor$ will yield an accurate representation of the radial problem.
Worse, simply increasing the order of the \cHp{} basis (i.e. increasing $p$) does not necessarily result in a more precise representation.
This may be understood as resulting from a basic limitation of approximation with fixed-order polynomial bases.
\par As a somewhat extreme but plausible example, consider the order-$50$ \cHp{} approximation of the number $1$,
\begin{align}
    \label{rep1}
    \ket{1} \; = \; \sum_{k=0}^{50} \, \ket{y_{50,k}} \brak{y_{50,k}}{1} \; .
\end{align}
Each polynomial, $\ket{y_{50,k}}$, has $51$ coefficients, and on the \rhs{} of \eqn{rep1}, values of $\brak{y_{50,k}}{1}$ must conspire such that only constant terms survive the sum (as is required since $1$ is a constant).
In practice, finite precision means that the cancellations will not be exact.
As a result, the net approximation will be contaminated by high frequency polynomial oscillations, or rather, \textit{polynomial noise}.
\par Clearly, in this case it would be wise to represent $\ket{1}$ with the order-$0$ \cHp{}.
This is consistent with the completeness of each \cHp{} fixed order basis: an order-$p$ \cHp{} basis can exactly represent polynomials of the same order; while lower order polynomials may also be represented, their representation in polynomial bases of much larger order increases the likelihood of numerical noise.
\par Further, for non-polynomial functions, it will not clear a priori which, if any, fixed order \cHp{} basis should be used.
For example, since the \qnm{s} radial functions, $\ket{f}$, are generally non-polynomial, one might attempt to apply a very large order \cHp{} representation,
\begin{align}
    \label{rep2}
    \ket{f} \; = \; \sum_{k=0}^{P} \, \ket{y_{Pk}} \brak{y_{Pk}}{f} \; ,
\end{align}
where $P \gg 1$.
In practice, the result would be contaminated by polynomial noise: all $\brak{y_{Pk}}{f}$ would have to be given to arbitrarily high precision such that the sum's $(P+1)^2$ polynomial coefficients (i.e. there are $P+1$ polynomials, and each with $P+1$ coefficients) add without significant contamination from round-off error.
\par In contrast, a basis of $P+1$ classical polynomials (i.e. with unique orders ranging from $0$ to $P$) would only have $\frac{1}{2}(P+1)^2$ coefficients to total. 
More importantly, unlike the order-$P$ \cHp{s}, such a basis would have expansion coefficients (analogs of $\brak{y_{Pk}}{f}$) that necessarily decrease as $k$ increases, meaning that fewer than $\frac{1}{2}(P+1)^2$ coefficients may ultimately be of practical importance.
\bpar{Towards canonical polynomials} Together, these ideas may be summarized as follows.
Since an order-$p$ \cHp{} basis is complete, it may be used to approximate the radial problem in a way that explicitly reduced to a polynomial solution in the limit that $\alpha_{p}=0$.
Further, it may be possible to treat general solutions of the radial problem perturbatively, where $\alpha_{p}=(p-\psp)(p-\psm)$ is the perturbative parameter.
However, since $\LcH$ will generally be non-band diagonal, the order-$p$ representation of \qnm{} radial functions, $\ket{f}$, may not quickly converge to a useful approximation.
Unlike approximation with classical polynomials, it is not always possible to reduce error by increasing the order of the \cHp{} basis used. 
\par These deficits are underpinned by the fact that each order-$p$ \cHp{} basis is a multiplex, i.e. all members of each \cHp{} basis are polynomials of the same order.
There is, therefore, good reason to wonder if there exists a basis of polynomials, each with a distinct order, that is orthogonal under the scalar product~(\ceqn{p14}).
\section{Natural \& Canonical Polynomials}
\label{s4}
\par The previous discussion does not rule out the \textit{construction} of polynomials that are \i{} complete in the infinite dimensional sense, \ii{} orthogonal \wrt the scalar product~(\ceqn{p14}), and \iii{} can represent $\mcD_\xi$ as a symmetric matrix.
While such polynomials cannot be exactly of the Heun type for the reasons stated in \sec{prelims}, they may yet possess the key features of classical polynomials, namely completeness and orthogonality.
In this sense, polynomials with properties \i{} and \ii{} will be referred to as \textit{canonical}.
Polynomials satisfying \iii{} will be referred to as \textit{natural} to $\mcD_\xi$.
\par For example, the Chebyshev polynomials defined on $\xi \in [0,1]$ are canonical, but not natural because $\mcD_\xi$ is not self-adjoint with respect to the Chebyshev inner-product. 
In contrast, the confluent Heun polynomials discussed in \secsa{prelims}{s3} are natural to $\mcD_\xi$, but due to their overcompleteness and atypical orthogonality, they are not canonical.
\par The possibility that the confluent Heun polynomials may be used to construct polynomials that are \textit{both} natural to $\mcD_\xi$, and canonical, is our present concern.
Henceforth, such polynomials will be referred to as \textit{\ccHp{s}}. 
When it is clear that confluent Heun polynomials are being discussed, the {\ccHp{s}} may be simply referred to as the ``canonical polynomials''.
\par In this section, two equivalent methods to construct \ccHp{s} are described.
The first method pertains to the construction of canonical polynomials directly from confluent Heun polynomials defined in \sec{prelims}.
This method will be referred to as ``\textit{direct construction}''.
The second method is based upon the standard \gs{} process.
This method will be referred to as ``\textit{\gs{} construction}''.
\par This section concludes with comments on the \textit{equivalence} between direct and \gs{} construction, their completeness, and finally two relevant \textit{representations} of the \ccHp{s} in terms of the \cHp{s}.
\bpar{Direct Construction} Let a \ccHp{} of order $p$ be denoted $\ket{u_p}$, where it is represented in the basis of order $p$ \cHp{s},
\begin{align}
    \label{p26}
    \ket{u_p} \; &= \; \sum_{k=0}^{p} \, \text{c}_{pk} \, \ket{{y}_{pk}} \;.
\end{align} 
The direct construction of $\ket{u_p}$ from $\ket{y_{pk}}$ requires the determination of constants, $\text{c}_{pk}$, such that the resulting polynomials are canonical (properties \ci{} and \cii{}), and natural to $\mcD_\xi$ (property \ciii{}).
We begin our construction with property \ii{}.
\par Explicitly, property \ii{} means that
\begin{align}
    \label{p27}
    \brak{u_{p'}}{u_{p}} \; \propto \; \delta_{p'p}\;.
\end{align}
We will now show that \eqnsa{p26}{p27} suffice to determine all expansion coefficients, $c_{pk}$, via an iterative linear process.
\par Without loss of generality, let $c_{pp}=1$.
It follows that, for $p=0$,
\begin{align}
    \label{p28}
    \ket{u_0}=\ket{y_{00}} \; .
\end{align}
For $p=1$, applying $c_{11}=1$ to \eqn{p26} means that $\ket{u_1}=c_{10}\ket{y_{10}}+\ket{y_{11}}$. 
Orthogonality~(\ceqn{p27}) then requires that 
\begin{align}
    \label{p29}
    \brak{u_0}{u_1} \; = \; c_{10}\brak{y_{00}}{y_{10}}+\brak{y_{00}}{y_{11}} \; = \; 0 \; .
\end{align}
Thus, the single unknown constant $c_{10}$ may be solved for, yielding 
\begin{align}
    \label{p30}
    c_{10} \; = \; -\frac{\brak{y_{00}}{y_{11}}}{\brak{y_{00}}{y_{10}}} \; .
\end{align}
\par For $p=2$, we may again consider $\ket{u_2}$ as defined in \eqn{p26}, namely $\ket{u_2}=c_{20}\ket{y_{20}}+c_{21}\ket{y_{21}}+\ket{y_{22}}$.
Here, $c_{20}$ and $c_{21}$ are the unknowns of interest.
Requiring that $\brak{u_0}{u_2}=0$ and $\brak{u_1}{u_2}=0$ results in two coupled equations,
\begin{subequations}
    \label{p31}
    \begin{align}
        \label{p31a}
        c_{20}\brak{u_0}{y_{20}}+c_{21}\brak{u_0}{y_{21}}+\brak{u_0}{y_{22}} \; &= \; 0 \; ,
        \\
        \label{p31b}
        c_{20}\brak{u_1}{y_{20}}+c_{21}\brak{u_1}{y_{21}}+\brak{u_1}{y_{22}} \; &= \; 0 \; .
    \end{align}
\end{subequations}
Moving the last terms in \eqnsa{p31a}{p31b} to their respective \rhs{s} yields that
\begin{subequations}
    \label{p32}
    \begin{align}
        \label{p32a}
        c_{20}\brak{u_0}{y_{20}}+c_{21}\brak{u_0}{y_{21}} \; &= \; -\brak{u_0}{y_{22}} \; ,
        \\
        \label{p32b}
        c_{20}\brak{u_1}{y_{20}}+c_{21}\brak{u_1}{y_{21}} \; &= \; -\brak{u_1}{y_{22}} \; .
    \end{align}
\end{subequations}
\Eqn{p32} is equivalent to the linear matrix equation,
\begin{align}
    \label{p33}
    \hat{X}_2 \; \vec{c}_2 \; = \; \vec{b}_2 \; ,
\end{align}
where $\hat{X}_2$ has elements ${X}_{2,jk}=\brak{u_j}{y_{2k}}$, $\vec{c}_2$ has elements $c_{2j}$, and $\vec{b}_2$ has elements $b_{2,j}=-\brak{u_j}{y_{22}}$.
\smallskip
\par In \eqnsa{p32}{p33}, the essential features of subsequent cases (i.e. $p>2$) are present:
For any $\ket{u_p}$, there will be $p$ equations of the form of \eqn{p32}, with a total of $p$ unknowns, and this system of linear equations will generally be equivalent to a linear matrix equation,
\begin{align}
    \label{p34}
    \hat{X}_p \; \vec{c}_p \; = \; \vec{b}_p \; .
\end{align}
As in \eqn{p34}, $\hat{X}_p$ and $\vec{b}_p$ will generally depend on all polynomials $\ket{u_{p'}}$ with $p'<p$, i.e. all previous iterations.
Since both $\{y_{pk}\}_{k=0}^{p}$ and $\{u_{p'}\}_{p'=0}^{p}$ are complete finite dimensional spaces, $\hat{X}_p$ plays the role of a change-of-basis matrix, allowing for the determination of $c_{pk}$ via
\begin{align}
    \label{p35}
    \vec{c}_p \; = \; \hat{X}_p^{-1} \, \vec{b}_p \; .
\end{align}
\par In \eqn{p35}, $\hat{X}_p$ and $\vec{b}_p$ are implicitly functions of all $\vec{c}_{p'}$, where $p'<p$.
Thus, given a solution for $\vec{c}_p$, one may then evaluate the \rhs{} of \eqn{p35} to compute $\vec{c}_{p+1}$ which in turn allows for the evaluation of the canonical polynomial, $\ket{u_{p+1}}$ via \eqn{p26}.
In this way, the general determination of \ccHp{s} follows inductively from \eqnsa{p26}{p27}.
\par Having constructed the polynomials, $u_p$, to be orthogonal, two scaling freedoms remain: signature and normalization.
\begin{subequations}
    \label{ccHp-norm}
    Since the scalar product~(\ceqn{p14}) is unchanged with respect to negation of both of its inputs (e.g. $\brak{u_p}{u_p}=\brak{-u_p}{-u_p}$), there is an overall $\pm1$ scaling, or rather ``signature'' freedom.
    Here, this is fixed by dividing each $u_p$ by its value at $\xi=1$,
    \begin{align}
        \label{ccHp-norm-a}
        {u_p} &\leftarrow {u_p}/u_p(1) \; .
    \end{align}
    The final freedom is that of normalization. Here, this is fixed in the usual way,
    \begin{align}
        \label{ccHp-norm-b}
        {u_p} &\leftarrow {u_p}/\sqrt{\brak{u_p}{u_p}} \; .
    \end{align}
\end{subequations}
\bpar{\gs{} Construction} The well-known \gs{} process acts on a linearly independent sequence, producing a new sequence that is orthogonal \wrt a chosen scalar product~\cite{Axler:2015}.
Given a predefined order for the original sequence, the resulting orthogonal sequence is known to be \textit{unique}, up to normalization choices. 
In this sense, the familiar classical polynomials (e.g. those of Hermite, Jacobi and Laguerre) are uniquely defined by applying \gs{} to the monomials, e.g. $\xi^p$, such that the resulting polynomial sequence is orthogonal \wrt the relevant inner product~\cite{abramowitz+stegun}.
\par Similarly, the \ccHp{s} may be defined by applying \gs{} to the monomials using the weight function, \eqn{p16}. 
Orthogonality is guaranteed because the algorithm iteratively removes the non-orthogonal components between different elements:
\begin{figure}[t]
    
    \hspace{-1cm}
    \begin{tabular}{c}
        \hspace{0pt}\includegraphics[width=0.44\textwidth]{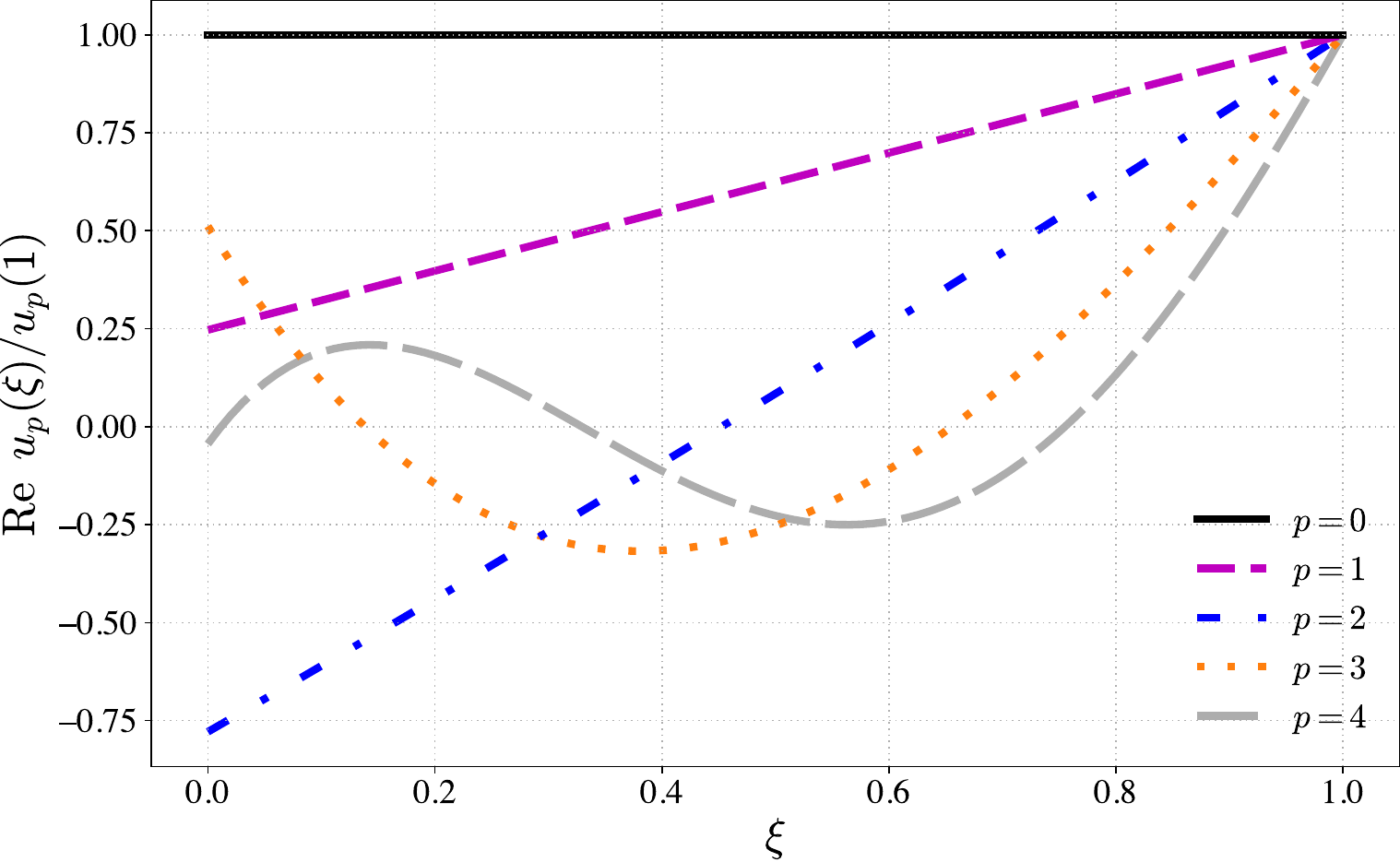}
        \vspace{4pt}
        \\
        \hspace{0pt}\includegraphics[width=0.44\textwidth]{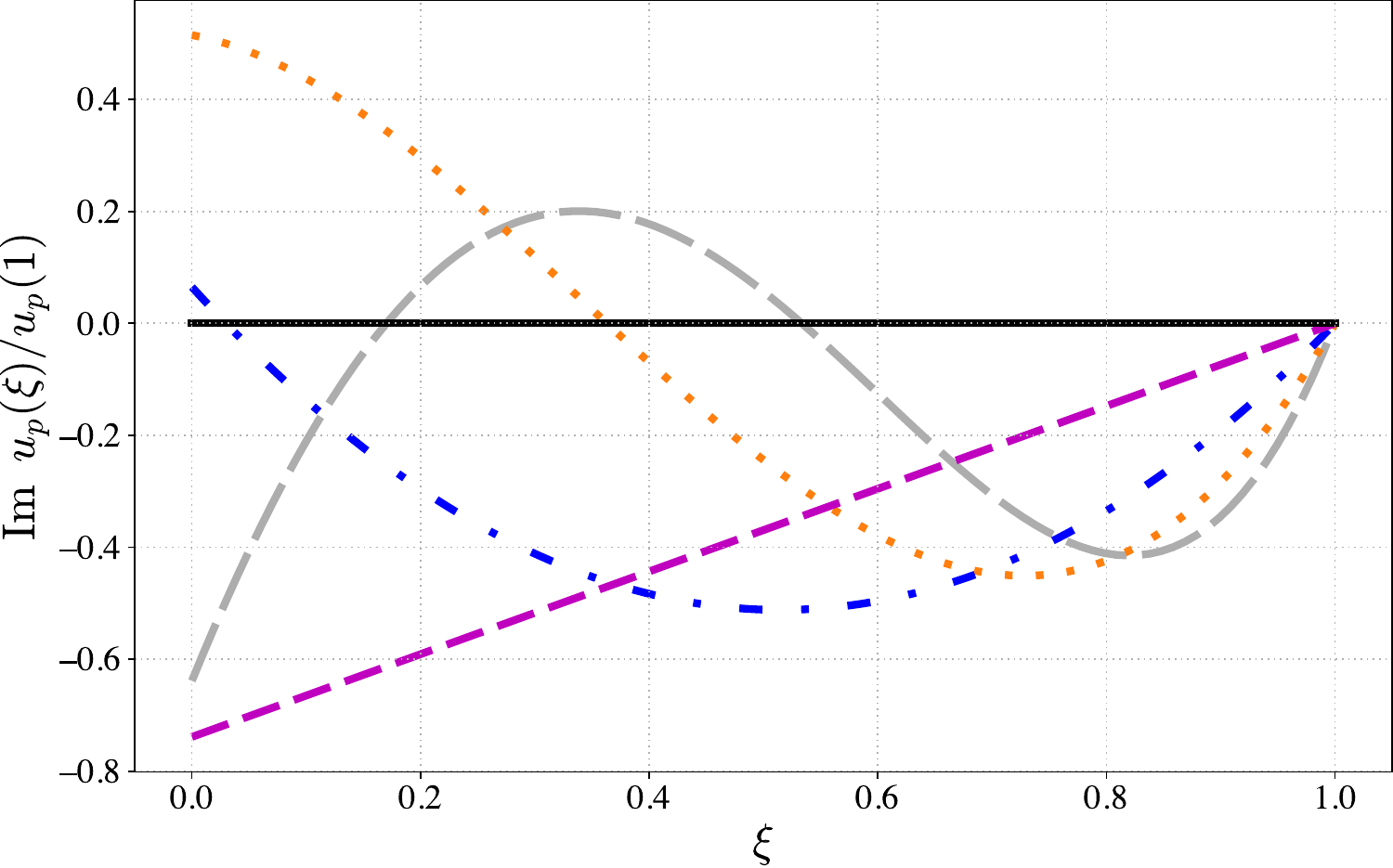}
    \end{tabular}

    \caption{ The first five \ccHp{s}, i.e. $u_p(\xi)$ with $p\in\{0,1,2,3,4\}$. 
    Polynomials are constructed using field spin weight $s=-2$, \bh{} spin $a/M=0.7$, and Kerr \qnm{} frequency $M\cw_{220}=0.5326-0.0808i$. 
    For ease of presentation, polynomials have been divided by their value at $\xi=1$ (i.e. spatial infinity). 
    Top panel: Real part of each polynomial. 
    Bottom panel: imaginary part of each polynomial. 
    Related discussion is provided in \sec{s6}. }
    \label{F2} 

\end{figure}
In particular, given a sequence of monomials $\{\xi^p \}_{p=0}^{\infty}$, the algorithm iteratively defines canonical polynomials, $u_{p}(\xi)$, to be $\xi^{p}$ plus a lower order polynomial that is constrained by requiring that $\brak{u_p}{u_{p'}}=0$ for all $p'<p$.
{The results is a sequence of orthogonal polynomials that can then be normalized according to \eqn{ccHp-norm}.}
\par It happens that that the weight function used to define the scalar product in \eqn{p14} is of the Pollaczek-Jacobi type, but with complex valued parameters~(\ceqnsa{p12}{p17}).
Since our weight function may generally be evaluated using analytic continuation~(see \ceqn{p20}), the \gs{} polynomials described above may be thought of as analytically continued Pollaczek-Jacobi polynomials~\cite{Chen:2010}.
Further connections between the polynomials described above, and the Pollaczek-Jacobi polynomials will be explored in future work~\cite{Foucoin:2024abc}. 
\bpar{Equivalence and Completeness} Both direct and \gs{} constructions of the \ccHp{s} are unique iterative processes with identical initial conditions, namely that $\ket{u_0}$ is a constant.
In each construction, the requirement of orthogonality, i.e. $\brak{u_{p'<p}}{u_p}=0$, uniquely defines $\ket{u_1}$, as well as subsequent polynomials.
Since, upon each iteration, there is only one non-trivial result for each $\ket{u_p}$, both processes result in normalized \ccHp{s} that are equivalent up to a root-of-unity scaling factor.
\par Equivalence is easily checked by e.g. writing $\ket{u_0}$ and $\ket{u_1}$ explicitly as polynomials in $\xi$, and then verifying that both constructions yield the same result, which is
\begin{subequations}
    \label{p38}
    \begin{align}
        \label{p38a}
        u_0(\xi) \; &= \; { \monm{0} }^{-1/2} \;,
        \\
        \label{p38b}
        u_1(\xi) \; &= \; \left[{\frac{ { \monm{0} } }{  { \monm{0}\monm{2}-\monm{1}^2 }  }}\right]^{1/2} \; \left( \xi-\frac{\monm{1}}{\monm{0}}   \right) \; .
    \end{align}
\end{subequations}
In \eqn{p38}, we recall that $\monm{k}$ are monomial moments defined by \eqn{p20}. 
On arrives at \eqnsa{p38a}{p38b} regardless of whether direct or \gs{} construction is used.
\par Henceforth, we will make no distinction between the \ccHp{s} generated by direct or \gs{} construction.
\par Regarding completeness, since the results of direct and \gs{} construction are equivalent, if the result of one construction is complete, then so is the other. 
The completeness of the \gs{} constructed polynomials is perhaps the most straightforward. 
Any sequence of \gs{} polynomials, if constructed from a complete set, will also be complete.
The monomials are complete when all monomial moments are well-defined\footnote{See Sec. VI of \PaperOne{} for a detailed discussion.} according to \eqn{p20}~\cite{London:202XP1,adkins2012algebra,Axler:2015}.
For example, in \PaperOne{}, it was found that the monomial moments are well-defined for all \qnm{} frequencies with non-zero real parts, which is the case for most \qnm{} frequencies of astrophysical relevance. 
Therefore, we expect that the \ccHp{s} will generally be complete in that setting.
\par Since the \cHp{s} are eigenfunctions to $\mcD{_\xi}$, it can be useful to represent each $\ket{u_p}$ in terms of \cHp{s}, $\ket{y_{p'k'}}$.
However, since the \cHp{s} are overcomplete, there is no unique representation.
\bpar{Single and Multi-order representations} The overcompleteness of the \cHp{s} means that there are {multiple equivalent representations} of {any} integrable function in terms of the \cHp{s}, $\ket{y_{p'k}}$.
Here, this fact is applied to the \ccHp{s}.
\par Since each space of order $p$ \cHp{s} is a $p+1$ dimensional basis, the identity operator for all order $p$ polynomials, $\I_p$, may be written as it was in \eqn{cHpI},
\begin{align}
    \label{p39}
    \I_p \; = \; \sum_{k=0}^{p} \, \ketbra{y_{pk}}{y_{pk}} \; .
\end{align}
Having defined the \ccHp{s}, $\ket{u_p}$, two mathematically different but practically equivalent representations result from \eqn{p39}. 
\par The first is simply the ansatz used for direct construction, i.e. \eqn{p26},
\begin{subequations}
    \label{p40}
    \begin{align}
        \label{p40a}
        \ket{u_p} \; &= \; \I_p \ket{u_p} \;
        \\
        \label{p40b}
                    &= \; \sum_{k=0}^{p} \ket{{y}_{pk}}\brak{y_{pk}}{u_p} \;
        \\
        \label{p40c}
                    &= \; \sum_{k=0}^{p} c_{pk}\,\ket{{y}_{pk}} \;.
    \end{align} 
\end{subequations}
In \eqnsa{p40a}{p40b}, the order-$p$ identity operator, $\I_p$, has been applied to $\ket{u_p}$.
In \eqn{p40c}, it is noted that $c_{pk}=\brak{y_{pk}}{u_p}$.
The direct construction algorithm discussed previously is simply a way to determine $\brak{y_{pk}}{u_p}$ when $\ket{u_p}$ is unknown.
\par Within \eqn{p40}, each \ccHp{} is represented in the basis of a different polynomial order, meaning that a set of multiple \ccHp{s} would require the generation of \cHp{s} from multiple orders.
\redoff{This is evident in \eqn{p40} by $\ket{u_p}$ on the \lhs{} having the same label, $p$, as the upper limit of the sum on the \rhs{}. Thus, for each different polynomial order, a different \cHp{} basis is considered.}
For this reason, \eqn{p40} will henceforth be referred to as the \textit{multi-order representation} of the \ccHp{s}. 
\par For any finite sequences of \ccHp{s}, $\{\ket{u_p}\}_{p=0}^{p'}$, the multi-order representation requires the evaluation of all polynomials within the set $\bigcup_{p=0}^{p'}\{\ket{y_{pk}}\}_{k=0}^{p}$.
Since there will be $\frac{1}{2}(p'+1)(p'+2)$ such polynomials, their evaluation presents a significant, but ultimately avoidable, computational expense.
\par A second representation allowed by \eqn{p39} entails choosing a single \cHp{} basis of order $p'$ to represent all \ccHp{s} with $p \le p'$,
\begin{subequations}
    \label{p41}
    \begin{align}
        \label{p41a}
        \ket{u_p} \; &= \; \I_{p'} \ket{u_p} \;
        \\
        \label{p41b}
                    &= \; \sum_{k=0}^{p'} \ket{{y}_{p'k}}\brak{y_{p'k}}{u_p} \;
        \\
        \label{p41c}
                    &= \; \sum_{k=0}^{p'} d_{p'pk} \, \ket{{y}_{p'k}} \;.
    \end{align} 
\end{subequations}
\redoff{We emphasize that, unlike \eqn{p40}, $\ket{u_p}$ in the \lhs{} of \eqn{p41} has a label $p$ that differs from the upper limit in the sum on the \rhs{} of \eqn{p41}.}
In \eqnsa{p41a}{p41b}, the identity operator for the order-$p'$ \cHp{s} has been applied to $\ket{u_p}$, and in \eqn{p41c},
\begin{align}
    d_{p'pk} \; = \; \brak{y_{p'k}}{u_p} \; .
\end{align}
\par In practice, one may first generate $\ket{u_p}$ via \gs{} construction, and then explicitly compute $\brak{y_{p'k}}{u_p}$ using \eqn{p19}.
\Eqn{p41} will henceforth be referred to as the \textit{single-order representation} of the \ccHp{s}.
If the largest polynomial order of interest is $p'$, then the single-order representation has the practical advantage of only requiring $p'+1$ \cHp{s} to be computed after \gs{} construction.
%
%
%
\section{Tridiagonalization}
\label{s5}
\par We may now return to \tk{}'s radial equation for \qnm{s},
\begin{align}
    \label{p42}
    \mcL{_\xi} \, \ket{f} \; &= \; A \, \ket{f} \; .
\end{align}
Having determined canonical polynomials that are natural to $\mcL{_\xi}$, their completeness and orthogonality may be used to represent \eqn{p42} as a simple linear matrix equation. 
\par Since the \ccHp{s} are complete, we may use them to represent the identity operator,
\begin{align}
    \label{p43}
    \I \; &= \; \sum_{p=0}^{\infty}\,\ket{u_p}\bra{u_p}  \; .
\end{align}
This allows each operator and vector within \eqn{p42} to be projected into the basis of \ccHp{s},
\begin{align}
    \label{p44}
    \I \,\mcL{_\xi} \,\I \, \ket{f} \; &= \; A \, \I \, \ket{f} \; .
\end{align}
Using \eqn{p43} to expand \eqn{p44} yields
\begin{subequations}
    \label{p45}
    \begin{align}
        \label{p45a}
        \sum_{p,q} \, \ket{u_{p}}&\bra{u_{p}} \, \mcL{_\xi} \,\ket{u_{q}} \brak{u_q}{f}\; \; 
        \\ \nonumber
        &= \; A \, \sum_{p} \, \ket{u_{p}}\brak{u_p}{f} \; ,
        \\
        \label{p45b}
        \hat{L}\,\vec{f} \; &= \; A\, \vec{f} \; .
    \end{align}
\end{subequations}
In going from \eqn{p45a} to \eqn{p45b}, we recognize $\hat{L}$ to be the matrix whose elements are
\begin{align}
    \label{p46}
    L_{pq} \; = \; \brak{u_{p}}{\mcL{_\xi} |\,u_{q}}  \; ,
\end{align}
and $\vec{f}$ to be the vector whose elements are 
\begin{align}
    \label{p47}
    f_p \; = \; \brak{u_p}{f} \; .
\end{align}
Recall that since $\mcL{_\xi}$ is self-adjoint \wrt the scalar product, the following equivalences hold: $\brak{u_{p}}{\mcL{_\xi} |\,u_{q}}  \; = \; \brak{u_{p}}{\mcL{_\xi} \,u_{q}}\; = \; \brak{\mcL{_\xi} \,u_{p}}{u_{q}}$.
\medskip
\par In this section, two mathematically different but practically equivalent routes to tridiagonalizing $\hat{L}$ are presented.
The first route is based upon the single-order representation of the \ccHp{s}, and it will be discussed in detail.
The second route is based upon their multi-order representation, and it will be described summarily as the step-by-step reasoning is both similar to that of the single-order approach, and already detailed in \PaperOne{}.
\par Regardless of the representation used, the result is the same: the confluent Heun operator $\mcL_\xi$ is approximated by a matrix, $\hat{L}$, that is of finite size, symmetric and tridiagonal.
In effect, the complex Jacobi matrix of $\mcL_\xi$ is determined~\cite{teschl2000jacobi,kuijlaars2003orthogonality}. 
\par Our present concern is the structure of $\hat{L}$.
Since its matrix elements are $\brak{u_{p}} { \mcL{_\xi} \,u_{q}}$, the action of $\mcL{_\xi}$ on $\ket{u_{q}}$ must first be understood, 
\begin{subequations}
    \label{p48}
    \begin{align}
        \label{p48a}
        \mcL{_\xi} \,\ket{u_{q}} \; = \; (\text{C}_0+\text{C}_1 [1-\xi] ) \,&\ket{u_{q}}
        \\
        \label{p48b}
        \; + \quad\quad\quad \;  \mathcal{D}_{\xi} \,&\ket{u_{q}} \; .
    \end{align}
\end{subequations}
In \eqn{p48}, as in \eqn{p11}, $\mcL_\xi$ has been written in terms of the physical constants $\tx{C}_0$ and $\tx{C}_1$~(\ceqn{p48a}), as well as the confluent Heun operator $\mcD_\xi$~(\ceqn{p48b}).
Plainly, \eqn{p48a} only contains constant rescalings of $\ket{u_p}$ and terms dependent on $\xi\ket{u_q}$, and \eqn{p48b} is simply $\mathcal{D}_{\xi} \,\ket{u_{q}}$.
Since $\brak{u_p}{u_q}=\delta_{pq}$, constant rescalings of $\ket{u_p}$ will only contribute to the diagonal of $\hat{L}$.
Therefore, to understand the structure of $\hat{L}$, we must first refine our understanding of $\xi\ket{u_q}$ and $\mathcal{D}_{\xi} \,\ket{u_{q}}$.
\smallskip
\par It happens that $\xi\ket{u_q}$ can be represented in terms of only three \cp{s},
\begin{align}
    \label{rec3}
    \xi \ket{u_q} = \sigma_{q0}\ket{u_{q-1}} + \sigma_{q1}\ket{u_{q}} + \sigma_{q2}\ket{u_{q+1}} \; .
\end{align}
\Eqn{rec3} is a three-term recursion relationship between \ccHp{s}.
It is a standard result in polynomial theory, and can be understood as resulting from completeness, orthogonality, and symmetry of the scalar product, namely $\brak{u_p}{\xi u_q}=\brak{\xi u_p}{u_q}$.
\par In \eqn{rec3}, the recursion coefficients, $\sigma_{q0}$, $\sigma_{q1}$ and $\sigma_{q2}$ are $p$-dependent complex numbers. 
Unlike many instances in polynomial theory where the recursion coefficients are known analytically, here they are to be treated numerically, as determined by evaluation of the scalar product:
\begin{subequations}
    \label{rec4}
    \begin{align}
        \label{rec4a}
        \sigma_{q0} \; &= \; \brak{u_{q-1}}{\xi  \,u_q} \;,
        \\
        \label{rec4b}
        \sigma_{q1} \; &= \; \brak{u_{q}}{\xi  \,u_q} \;,
        \\
        \label{rec4c}
        \sigma_{q2} \; &= \; \brak{u_{q+1}}{\xi  \,u_q} \;.
    \end{align}
\end{subequations}
In the context of $\mcL{_\xi} \,\ket{u_{q}}$~(i.e. \ceqn{p48}), the existence of a three-term recursion relationship between polynomials means that terms within \eqn{p48a} only contribute to the central, lower and upper diagonal of $\hat{L}$.
Therefore, \eqn{p48a} will generally contribute to the tridiagonal part of $\hat{L}$.
\par This should be expected of \eqn{p48a}, since all canonical type polynomials (e.g. Chebyshev and Laguerre) have three-term recursions.
However, since not all \cp{s} are natural to $\mcD_\xi$, it happens that $\mcD_\xi \ket{u_q}$ is a problem-specific quantity.
\par Further, since $\mcD_\xi$ contains derivatives, it is here that the representation of each $\ket{u_q}$ has meaningful effect: while the previously discussed single- and multi-order representations are equivalent~(\ceqnsa{p40}{p41}), their behaviors under differentiation are not. 
\bpar{The single-order approach} We may first consider the single-order representation of the \ccHp{s}.
In that picture, {any} polynomial of order $q$ is expanded in terms of \cHp{s} of order $p' \ge q$ (\ceqn{p42}).
For a \ccHp{} $\ket{u_q}$, this expansion is written as 
\begin{align}
    \label{p49}
    \ket{u_q} \; &= \; \sum_{k=0}^{p'} d_{p'qk} \, \ket{{y}_{p'k}} \;.
\end{align} 
Therefore the action of $\mcD{_\xi}$ on $\ket{u_q}$ is given by,
\begin{subequations}
    \label{p50}
    \begin{align}
        \label{p50a}
        \mcD{_\xi}\,\ket{u_q} \; &= \; \sum_{k=0}^{p'} d_{p'qk} \, \mcD{_\xi} \, \ket{{y}_{p'k}} \;
        \\ 
        \label{p50b}
        &= \; \sum_{k=0}^{p'} \, d_{p'qk} \, (\lambda_{p'k}\,+\,\mu_{p'}\,\xi) \, \ket{{y}_{p'k}} \; ,
        \\ 
        \label{p50c}
        &= \; \ket{v_{p'q}} \, + \, \mu_{p'} \,\xi \,\ket{u_q} \; .
    \end{align} 
\end{subequations}
In \eqn{p50b}, the two parameter eigenvalue relation for the \cHp{s}, \eqn{p22a}, has been applied.
In \eqn{p50c}, terms that depend on $\lambda_{p'k}$ have been grouped to define a new polynomial,
\begin{align}
    \label{p51}
    \ket{v_{p'q}} \; = \; \sum_{k=0}^{p'} \, {d}_{p'qk} \, \lambda_{p'k} \, \ket{{y}_{p'k}} \; ,
\end{align}
and it has been noticed that remaining terms in \eqn{p50b} are simply proportional to $\ket{u_q}$.
Since $\mcD_\xi$ increases the order of polynomials by one~(see \csec{prelims}), $\ket{v_{p'q}}$ has order $q+1$.
\par It follows (from \ceqnsa{p48}{p50c}) that $\mcL_\xi\,\ket{u_q}$ may be written as 
\begin{subequations}
    \label{p52}
    \begin{align}
        \label{p52a}
        \mcL{_\xi} \ket{u_{q}} \; =& \; (\text{C}_0+\text{C}_1)\ket{u_{q}} 
        \\
        \label{p52b}
        &\;+ \; \ket{v_{p'q}}  
        \\
        \label{p52c}
        &\;+ \; \alpha_{p'} \, \xi \, \ket{u_q} .
    \end{align}
\end{subequations}
In \eqn{p52c}, recall that $\alpha_{p'}$ is defined in \eqn{alpha-p-factored}.
The structure of \eqn{p52} enables a few useful observations.
\Eqns{p52a}{p52c} imply that $\hat{L}$ is the sum of three matrices, $\hat{L}^{(\tx{S})}_0$, $\hat{L}^{(\tx{S})}_1$, and $\hat{L}^{(\tx{S})}_2$,
\begin{align}
    \label{p53}
    \hat{L} \; = \; \hat{L}^{(\tx{S})}_0 + \hat{L}^{(\tx{S})}_1 + \hat{L}^{(\tx{S})}_2 \; .
\end{align}
Their respective matrix elements are
\begin{subequations}
    \label{p54}
    \begin{align}
        \label{p54a}
        {L}_{0,pq}^{(\tx{S})}  \; &= \; (\text{C}_0+\text{C}_1) \, \delta_{pq}\;,
        \\
        \label{p54b}
        {L}_{1,pq}^{(\tx{S})}  \; &= \; \brak{u_p}{v_{p'q}} \;,
        \\
        \label{p54c}
        {L}_{2,pq}^{(\tx{S})}  \; &= \; \alpha_{p'} \braket{u_p}{\xi \, u_q}\;.
    \end{align}
\end{subequations}
Since $\mcL_\xi$ is self-adjoint \wrt the scalar product, $\hat{L}$ will be symmetric, meaning that $L_{pq}=L_{qp}$.
It may be observed that $\hat{L}^{(\tx{S})}_0$ and $\hat{L}^{(\tx{S})}_2$ are symmetric, owing to the symmetry of the Kronecker delta, and of the scalar product.
Thus for $\hat{L}^{(\tx{S})}_0 + \hat{L}^{(\tx{S})}_1 + \hat{L}^{(\tx{S})}_2$ to be symmetric, $\hat{L}^{(\tx{S})}_1$ must be symmetric, meaning that
\begin{align}
    \label{p55}
    \braket{u_p}{v_{p'q}} \; = \; \brak{v_{p'p}}{u_q}  \; .
\end{align}
This conclusion may also be reached via the consideration of $\brak{u_p}{\mcD{_\xi} \,u_{q}} \;= \;\brak{\mcD{_\xi} \,u_p}{u_{q}}$, which directly reduces to \eqn{p55}.
Thus, \eqn{p55} is only true for scalar products under which $\mcD_\xi$ is self-adjoint (e.g. \ceqn{p55} does not hold for Chebyshev inner-product, meaning that the associated $\hat{L}^{(\tx{S})}_1$ matrix would not be symmetric)~\cite{London:202XP1}.
\par Lastly, it may be deduced that $\hat{L}^{(\tx{S})}_1$ is tridiagonal.
This follows from the fact that each $\ket{v_{p'q}}$ is of order $q+1$.
In particular, given any order $q+1$ polynomial, e.g. $\ket{v_{p'q}}$, its \ccHp{} representation will not contain contributions from $\ket{u_p}$ if $p>q+1$,
\begin{subequations}
    \label{p56}
    \begin{align}
        \label{p56a}
        \ket{v_{p'q}} \; =& \; \I \,  \ket{v_{p'q}} 
        \\
        \label{p56b}
        =& \; \sum_{p=0}^{\infty} \, \ket{u_{p}}\brak{u_p}{v_{p'q}} \; ,
        \\
        \label{p56c}
        =& \; \sum_{p=0}^{q+1} \, \ket{u_{p}}\brak{u_p}{v_{p'q}} \; .
    \end{align}
\end{subequations}
In \eqns{p56a}{p56b}, a \ccHp{} representation is applied to $\ket{v_{p'q}}$.
In \eqn{p56c}, it has been acknowledged that terms with $p>q+1$ do not contribute; otherwise, the resulting polynomial would be of order greater than $q+1$.
This condition on $\brak{u_p}{v_{p'q}} $, as well as its counterpart on $\brak{u_q}{v_{p'p}} $ may be stated thusly,
\begin{subequations}
    \label{p57}
    \begin{align}
        \label{p57a}
        \brak{u_p}{v_{p'q}} \; = \; 0 \text{,  if }\;\; p>q+1\; ,
        \\
        \label{p57b}
        \brak{u_q}{v_{p'p}} \; = \; 0 \text{,  if }\;\; q>p+1\; .
    \end{align}
\end{subequations}
Note that $\brak{u_q}{v_{p'p}}=\brak{v_{p'p}}{u_q}$, not due to \eqn{p55}, but rather to the multiplicative associative property applied to the integrand of the scalar product~(\ceqn{p14}).
Also note that the conditional in \eqn{p57b} is equivalent to $p<q-1$.
\par Since \eqnsa{p57a}{p57b} hold simultaneously, it follows that $\brak{u_p}{v_{p'q}}\neq0$ when $q-1 \le p \le q+1$. 
In other words, each $\ket{v_{p'q}}$ is exactly equal to a sum over only three \ccHp{s},
\begin{align}
    \label{p58}
    \ket{v_{p'q}} \; = \; \nu_{q0}\ket{u_{q-1}} + \nu_{q1}\ket{u_{q}} + \nu_{q2}\ket{u_{q+1}} \; .
\end{align}
In \eqn{p58}, $\nu_{q0}$, $\nu_{q1}$ and $\nu_{q2}$ are complex valued $q$-dependent numbers defined by the scalar product,
\begin{subequations}
    \label{p59}
    \begin{align}
        \label{p59a}
        \nu_{q0} \; &= \; \brak{u_{q-1}}{v_{p'q}} \;,
        \\
        \label{p59b}
        \nu_{q1} \; &= \; \brak{u_{q}}{v_{p'q}} \;,
        \\
        \label{p59c}
        \nu_{q2} \; &= \; \brak{u_{q+1}}{v_{p'q}} \;.
    \end{align}
\end{subequations}
The elements of $\hat{L}^{(\tx{S})}_1$, namely $\brak{u_p}{v_{p'q}}$, are only nonzero between its upper and lower diagonal, so $\hat{L}^{(\tx{S})}_1$ is a tridiagonal matrix. 
\par Since $\hat{L}^{(\tx{S})}_1$ and $\hat{L}^{(\tx{S})}_2$ are symmetric and tridiagonal, and $\hat{L}^{(\tx{S})}_0$ is diagonal, if follows that $\hat{L}=\hat{L}^{(\tx{S})}_0 + \hat{L}^{(\tx{S})}_1 + \hat{L}^{(\tx{S})}_2$ is symmetric and tridiagonal.
Equivalently, 
\begin{align}
    \label{p60}
    \mcL_\xi \ket{u_q}  \; = \;\nu_{q3} \, \ket{u_{q-1}} 
    + \nu_{q4} \, \ket{u_{q}} 
    + \nu_{q5} \, \ket{u_{q+1}} \; ,
\end{align} 
where, 
\begin{subequations}
    \label{p61}
    \begin{align}
        \nu_{q3} \; &= \; \alpha_{p'}\,\sigma_{q0} + \nu_{q0} \; ,
        \\
        \nu_{q4} \; &= \; \alpha_{p'}\,\sigma_{q1} + \nu_{q1} + (\text{C}_0+\text{C}_1) \; ,
        \\
        \nu_{q5} \; &= \; \alpha_{p'}\,\sigma_{q2} + \nu_{q2} \; .
    \end{align}
\end{subequations}
In \eqn{p61}, we recall from \eqn{alpha-p-factored} that $\alpha_{p'}  =  \mu_{p'}-\tx{C}_1$, and that $\sigma_{q0}$, $\sigma_{q1}$ and $\sigma_{q2}$ are defined by \eqn{rec4}.
\bpar{The multi-order approach}
Here, an alternative approach to tridiagonalization is summarized, following and building upon Sec.~VIII of \PaperOne{}.
This approach uses the multi-order representation of the \ccHp{s}, 
\begin{align}
    \label{p62}
    \ket{u_q} \; &= \; \sum_{k=0}^{q} c_{qk} \, \ket{{y}_{qk}} \;.
\end{align} 
While the net result, namely $\hat{L}$, is the same as that found previously, the multi-order approach enables the derivation of relationships between $\hat{L}$ and the \ccHp{s'} three-term recursion that are ostensibly hidden within the single-order approach. 
This point is illustrated by first applying the multi-order representation of the \ccHp{s} to the radial problem, and then using the symmetry of the Jacobi matrix $\hat{L}$ to relate elements of $\hat{L}$ to the \ccHp{s'} three-term recursion.
This proceeds as follows.
\par Under the multi-order approach, $\hat{L}$ may be written as the sum of three matrices,
\begin{align}
    \label{p63}
    \hat{L} \; = \; \hat{L}^{(\tx{M})}_0 + \hat{L}^{(\tx{M})}_1 + \hat{L}^{(\tx{M})}_2 \; .
\end{align}
Their respective matrix elements are
\begin{subequations}
    \label{p64}
    \begin{align}
        \label{p64a}
        {L}_{0,pq}^{(\tx{M})}  \; &= \; (\text{C}_0+\text{C}_1) \, \delta_{pq}\;,
        \\
        \label{p64b}
        {L}_{1,pq}^{(\tx{M})}  \; &= \; \brak{u_p}{v_{q}} \;,
        \\
        \label{p64c}
        {L}_{2,pq}^{(\tx{M})}  \; &= \; \alpha_q \, \braket{u_p}{\xi \, u_q}\;.
    \end{align}
\end{subequations}
In \eqn{p64b}, $\ket{v_q}$ is an order-$q$ ``auxiliary'' polynomial defined as
\begin{align}
    \label{p65}
    \ket{v_q} \; = \; \sum_{k=0}^{q} \, \text{c}_{qk} \, \lambda_{qk} \, \ket{{y}_{qk}} \; ,
\end{align}
where it is recalled that $\lambda_{qk}$ is the eigenvalue of the \cHp{}, $\ket{y_{qk}}$.
Consideration of $\brak{u_p}{\mcD{_\xi} \,u_{q}} \;= \;\brak{\mcD{_\xi} \,u_p}{u_{q}}$ yields a that $\ket{v_q}$ may be equated with a sum over only a few \ccHp{s},
\begin{align}
    \label{p66}
    \ket{v_q} = \sigma_{q3}\ket{u_{q-1}} + \sigma_{q4}\ket{u_{q}}  \; .
\end{align}
In \eqn{p66}, $\sigma_{q3}$ and $\sigma_{q4}$ are complex valued $q$-dependent numbers defined by the scalar product,
\begin{subequations}
    \label{p67}
    \begin{align}
        \label{p67a}
        \sigma_{q3} \; &= \; \brak{u_{q-1}}{v_{q}} \;,
        \\
        \label{p67b}
        \sigma_{q4} \; &= \; \brak{u_{q}}{v_{q}} \;.
    \end{align}
\end{subequations}
It may be observed that $\hat{L}^{(\tx{M})}_0$ is diagonal, $\hat{L}^{(\tx{M})}_1$ upper-tridiagonal~(due to \ceqn{p66}), and $\hat{L}^{(\tx{M})}_2$ is tridiagonal.
Therefore, $\hat{L}$ is tridiagonal as expected.
\par It is informative to note that, while  $\hat{L}^{(\tx{M})}_0$ is symmetric, \eqnsa{p64b}{p64c} are \textit{asymmetric}:
\eqn{p64b} is asymmetric due to \eqn{p66} not having a contribution from $\ket{u_{q+1}}$ (i.e. $\hat{L}^{(\tx{M})}_1$ is only upper-tridiagonal), and \eqn{p64c} is asymmetric due to the $q$-dependent scale factor, $\alpha_q$.
Thus, since $\hat{L}$ is symmetric, $ \hat{L}^{(\tx{M})}_1$ and $ \hat{L}^{(\tx{M})}_2$ must combine to form a quantity that is also symmetric.
In turn, this means that $ \hat{L}^{(\tx{M})}_1$ and $ \hat{L}^{(\tx{M})}_2$ are not independent objects.
\par The link between $ \hat{L}^{(\tx{M})}_1$ and $ \hat{L}^{(\tx{M})}_2$ is perhaps easiest to understand by noting that the multi-order representation of the \ccHp{s} allows $\mcL_\xi \ket{u_q}$ to be written as,
\begin{align}
    \label{p68}
    \mcL_\xi \ket{u_q}  \; = \;\sigma_{q5} \, \ket{u_{q-1}} 
    + \sigma_{q6} \, \ket{u_{q}} 
    + \sigma_{q7} \, \ket{u_{q+1}} \; ,
\end{align} 
where
\begin{subequations}
    \label{p69}
    \begin{align}
        \label{p69a}
        \sigma_{q5} \; &= \; \alpha_q \, \sigma_{q0} + \sigma_{q3} \; ,
        \\
        \label{p69b}
        \sigma_{q6} \; &= \; \alpha_q \, \sigma_{q1} + \sigma_{q4} + (\text{C}_0+\text{C}_1) \; ,
        \\
        \label{p69c}
        \sigma_{q7} \; &= \; \alpha_q \, \sigma_{q2} \; .
    \end{align}
\end{subequations}
Plainly stated, since $\hat{L}$ is a symmetric matrix its matrix elements, $L_{pq}$, are such that $L_{pq}=L_{qp}$.
\Eqn{p69} means that $L_{pq}$ will be within the set $\{\sigma_{q5},\sigma_{q6},\sigma_{q7}\}$ for $p\in \{q-1,q,q+1\}$, respectively.
Since $\hat{L}$ is tridiagonal, the only nontrivial symmetry statement is $L_{q-1,q}=L_{q,q-1}$, which is equivalent to $L_{q,q+1}=L_{q+1,q}$ under $q\rightarrow q+1$.
Using equation \eqnsa{p68}{p69}, $L_{q-1,q}=L_{q,q-1}$ becomes,
\begin{subequations}
    \label{p70}
    \begin{align}
        \label{p70a}
        \sigma_{q5} \; &= \; \sigma_{q-1,7} \; ,
        \\
        \label{p70b}
        \alpha_q \, \sigma_{q0} + \sigma_{q3} \; &= \;  \; \alpha_{q-1}\,  \sigma_{q-1,2} \; .
    \end{align}
\end{subequations}
In \eqn{p70a}, $L_{q-1,q}=L_{q,q-1}$ has been rewritten using \eqn{p68}. 
\begin{figure*}[t]
    
    
    \includegraphics[width=\textwidth]{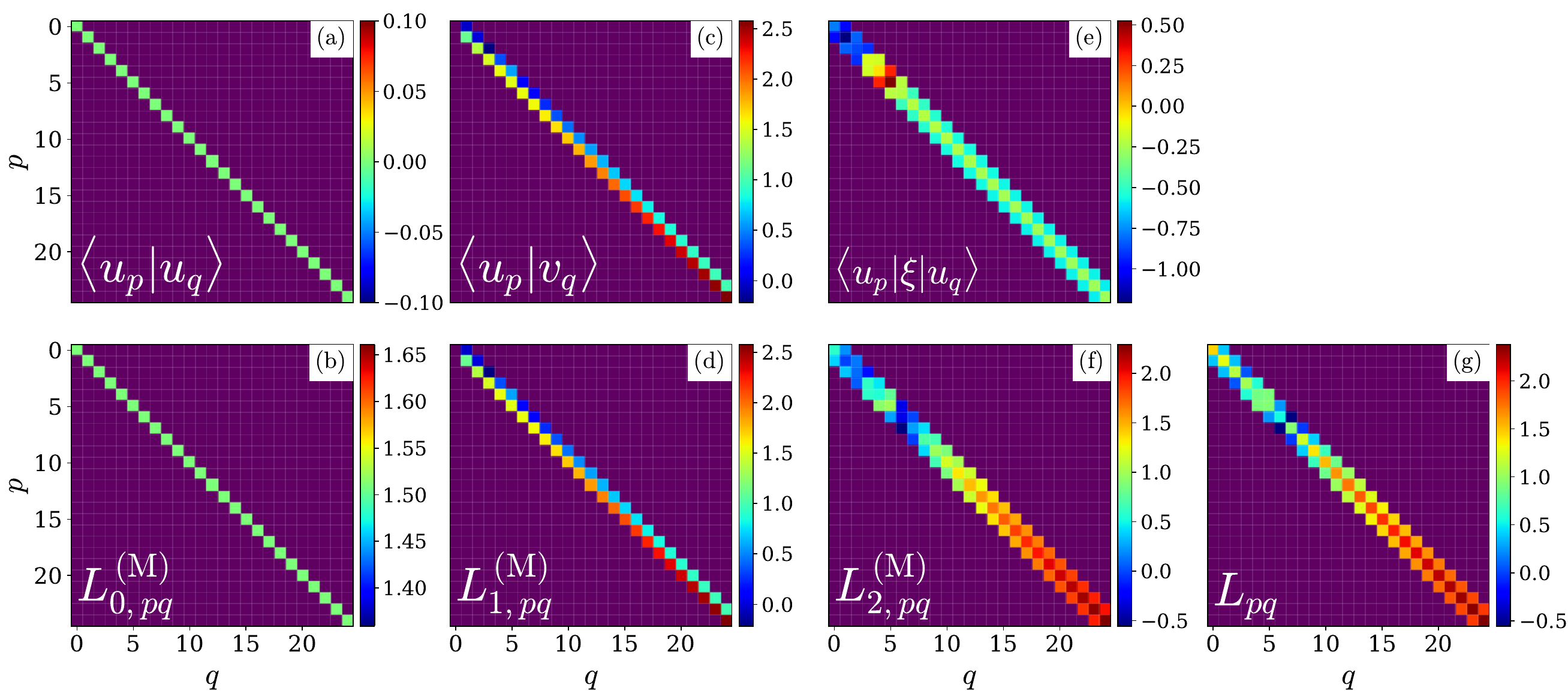} 
        
    \caption{ 
        Matrices related to \ccHp{s}, and representations of \tk{'s} radial operator for $(s,\ell,m,n)=(-2,2,2,6)$, $a/M=0.7$, and $M\cw_\lmn=0.4239-1.0954i$. All color bars correspond to the $\log_{10}$ of the absolute values of quantities noted in annotation, e.g.  for panel (a), $\log_{10}|\brak{u_p}{u_q}|$ is visualized. 
        Tiles (in purple) away from each panel's tridiagonal band are exactly zero. 
        See \sec{s6} for related discussion.
      }
    \label{F3}
\end{figure*}
In going from \eqn{p70a} to \eqn{p70b}, \eqn{p69} has been used to expand both sides.
\Eqn{p70b} may be further reduced by noting that symmetry of the canonical polynomials' three-term recursion~(See \ceqn{rec3}) means that $\sigma_{q-1,2} = \sigma_{q0}$.
Therefore, \eqn{p70b} allows $\sigma_{q3}$ (i.e. $\brak{u_{q-1}}{v_{q}}$) to be written as,
\begin{subequations}
    \label{p71}
    \begin{align}
        \label{p71a}
        \sigma_{q3} \; &= \;   (\alpha_{q-1}-\alpha_q) \, \sigma_{q0} \; ,
        \\
        \label{p71b}
        \brak{u_{q-1}}{v_{q}} \; &= \;  (2(q-1)+\tx{C}_4) \, \brak{u_{q-1}}{\xi  \,u_q} \; . 
    \end{align}
\end{subequations}
In \eqn{p71a}, \eqn{p70b} as been solved for $\sigma_{q3}$.
In \eqn{p71b}, $\sigma_{q3}$ and $\sigma_{q0}$ have been replaced with their scalar-product definitions via \eqnsa{p48a}{p67a}, and \eqnsa{p22b}{alpha-p} have been used to rewrite instances of $\alpha_q$.
\par \Eqn{p71b} communicates that the non-zero off-diagonal elements of $\hat{L}^{(\tx{M})}_1$, namely $\brak{u_{q-1}}{v_{q}}$, are directly proportional to the three-term-recursion coefficient, $\brak{u_{q-1}}{\xi  \,u_q}$.
Recalling \eqn{p64c}, the same is clearly true for off-diagonal elements of $\hat{L}^{(\tx{M})}_2$.
Together, those two observations mean that the off-diagonal elements of $\hat{L}$ are generally proportional to the three-term-recursion coefficients $\sigma_{q0}$ and $\sigma_{q2}$.
%
\section{Select numerical results}
\label{s6}
Before being put into practice, the preceding theoretical development benefits from a few points of guidance~\cite{London:202XP1}.
For all computations, physical parameters, namely $\{a/M,s,\cw,m\}$ must be chosen such that the derived parameters, $\COne$ through $\CFour$, may then be computed according to \eqn{p12}.
For nearly all subsequent manipulations involving polynomials or series expansions, the fundamental operation is the scalar product~(See \ceqn{p14}).
\par While it is possible to evaluate the scalar product by directly integrating it, success in that endeavor requires the formulation of a complex path that avoids singularities in the integrand~\cite{London:202XP1}. 
A much simpler method of evaluating the scalar product is given by \eqnsa{p19}{p20}, which reduces the effort to the evaluation of standard special functions via their well-known analytic continuations. 
\par Given the scalar product, it is subsequently useful to determine the eigenvectors and values of matrices, such as $\hat{L}$ which encodes the \qnm{} problem~(\ceqn{p45b}).
This may be accomplished with any of several standard linear algebra packages~\cite{2020NumPy-Array,mpmath}. 
Henceforth, \eqn{p19} is used to compute all scalar product, and \eqn{p20} is implemented with $128$ decimal points of general precision arithmetic via the \texttt{Python} package, \texttt{mpmath}~\cite{mpmath}.
This package is also used to solve various matrix eigenvalue problems.
\par In this section, example applications of the \ccHp{s} are provided along with related practical guidance.
While many cases are considered, most examples are connected by a single fiducial case: a Kerr \bh{} of dimensionless spin $a/M=0.7$, and its \qnm{} with $(s,\ell,m,n)=(-2,2,2,6)$.
Broadly, examples range from the brief consideration of the \ccHp{s}, to their application to the \cHp{} and \qnm{} problems. 
In the latter example, the \qnm{s'} radial functions are explicitly computed and shown to satisfy \tk{'s} radial equation~(\ceqn{p13a}).
This section concludes with a brief discussion of the transition from Schwarzschild to Kerr, from the perspective of the \qnm{} radial functions.
%
%
\bpar{\CcHp{s}}
Examples of the first five \ccHp{s} are shown in \fig{F2}.
\par As seen in the top panel of \fig{F2}, the polynomials may be formatted such that their real parts have zero roots along the real line, meaning that the net complex valued polynomials will have no roots at real values of the fractional radial coordinate, $\xi$.
This is simply a result of the fact that the \ccHp{s} have complex valued coefficients~(\ceqn{p38}), and so will generally have roots situated away from the real line.
%
%
\bpar{Tridiagonalization}
The \ccHp{s} may be used to tridiagonalize confluent Heun equations such as the \qnm{'s} radial equation~(\ceqn{p13}), or the \cHp{} equation~(\ceqn{p22}).
\par In \fig{F3} a step-by-step application of the \ccHp{s} to the \qnm{} radial problem is shown. 
There, the top three panels (i.e. \hyperref[F3]{$(a)$}, \hyperref[F3]{$(c)$} and \hyperref[F3]{$(e)$}) show quantities derived directly from the \ccHp{s}.
The remaining panels show how the multi-order representation~(\ceqn{p62}) of each quantity contributes to the representation of the radial differential operator according to \eqn{p63}.
\par When written in terms of matrix elements, \eqn{p63} is
\begin{align}
    \label{p63-2}
    L_{pq} \; = \; {L}^{(\tx{M})}_{0,pq} + {L}^{(\tx{M})}_{1,pq} + {L}^{(\tx{M})}_{2,pq} \; .
\end{align}
Under this framing, the qualitative features of panels \hyperref[F3]{$(b)$}, \hyperref[F3]{$(d)$} and \hyperref[F3]{$(f)$} are determined by those of panels \hyperref[F3]{$(a)$}, \hyperref[F3]{$(c)$} and \hyperref[F3]{$(e)$}, respectively. 
\par Conceptually, \ccHp{} orthogonality~(\ceqn{p27}) is demonstrated in panel \hyperref[F3]{$(a)$}, upper-tridiagonality of auxiliary polynomials~(\ceqn{p66}) is demonstrated in panel \hyperref[F3]{$(c)$}, and three-term recursion~(\ceqn{rec4}) is dempstrated in panel \hyperref[F3]{$(e)$}.
Then end result, panel \hyperref[F3]{$(g)$}, may be considered to represent the three-term recursion of orthogonal polynomials which approximate solutions to the \qnm{} radial problem~\cite{teschl2000jacobi,kuijlaars2003orthogonality}. 
%
%
\redoff{\bpar{Matrix asymptotics \& truncation} 
Even if the operator $\mcL_\xi$ is self-adjoint with respect to a given scalar product, this does not guarantee that its matrix representation will have eigenvalues that accurately approximate those of $\mcL_\xi$.
The extent of agreement between the two sets of eigenvalues depends on the asymptotic behavior of the matrix’s off-diagonal elements~\cite{Petropoulou:2014eug}.
%
%
By asymptotic, we refer to the behavior of matrix elements, $L_{pq}$ with $q\in\{p-1,p,p+1\}$, as $p$ tends towards infinity.
\par For example, it is well known that the spherical harmonics are an excellent basis in which to represent the \qnm{s}' spheroidal harmonics~\cite{OSullivan:2014ywd,Teukolsky:1973ha,Cook:2014cta,Fackerell:1977,London:2020uva}.
In part, this is because, when represented in a spherical harmonic basis, the spheroidal harmonic differential operator is asymptotically diagonal\footnote{\redoff{This can be seen computing the spherical-spheroidal mixing coefficients, and then considering the limit as $\ell\rightarrow\infty$}~\cite{London:2020uva,Berti:2014fga}.}, meaning that large-$\ell$ spheroidal harmonics are asymptotically equivalent spherical harmonics of the same $\ell$~\cite{London:2020uva,Fackerell:1977,Cook:2014cta}. As a result, one can typically estimate spheroidal harmonic eigenvalues with little concern for truncation error by first casting the problem into a spherical harmonic basis, and then solving the corresponding matrix eigenvalue problem~\cite{Cook:2014cta,Fackerell:1977,London:2020uva,OSullivan:2014ywd}. 
\smallskip
\par We now discuss why this is typically not the case for the \qnm{s'} radial problem. In short, off-diagonal matrix elements for the radial problem typically grow \textit{at least as fast} as the diagonal ones, meaning that the matrix does not become asymptotically diagonal. 
This behavior is explicit in the radial functions' Frobenius-type series solution as studied by Refs.~\cite{leaver85,Nollert:1999ji,Leaver86c,Cook:2014cta}, and is therefore necessarily present in any orthonormal polynomial representation of the same problem\footnote{\redoff{This is due to the fact that the 3-term recursions for the series solution must be asymptotically equivalent to the effective tridiagonalization of any polynomial representation. While many polynomial representations will not result in tridiagonal operators, all such operators can be numerically tridiagonalized using standard techniques}~\cite{2020SciPy-NMeth}.}~\cite{Chung:2023wkd,Jaramillo:2020tuu,Dias:2013hn}.  
In effect, the slow convergence of the radial functions' series solutions, along with truncation thereof, results in what is known as \textit{spectral pollution}~\cite{Davies2003math2145D,Lewin2008arXiv0812.2153L,BOULTON20161,MartnezAdame2007}.
%
\par For the \ccHp{s}, panel \hyperref[F3]{$(e)$} of \fig{F3} illustrates a general observation, namely that the large-$p$ structure of 3-term recursion is that of a tridiagonal Toeplitz matrix (i.e. a symmetric matrix with constant elements along its diagonals). 
It follows from the role of $\alpha_{p} \; = \; (p-\psp)(p-\psm)$ in \eqn{p69} that all $L_{pq}$ (with $q\in\{p-1,p,p+1\}$) are asymptotically equivalent to $p^2$.
Therefore, the matrix representation of the radial operator is not asymptotically diagonal, and not self-adjoint in the sense defined in Ref.~\cite{Petropoulou:2014eug}.
The related qualitative behavior is seen in panel \hyperref[F3]{$(g)$} of \fig{F3}, which shows matrix elements having a local minimum near $p=q=6$, before becoming increasingly comparable as $p$ and $q$ increase.
\par In forthcoming numerical results, we show both the eigenvalues that result from spectral pollution, as well as those consistent with the non-truncated (infinite dimensional) eigenvalue problem. 
In the latter case, eigenvalues were determined by using an $N=60$ dimensional \ccHp{} basis, and then truncation-induced eigenvectors were removed. 
Consistency with the non-truncated problem was checked by comparing against the
continued-fraction result\footnote{\redoff{The continued-fraction result is also prone to spectral pollution. However, it can typically be avoided in that context, if $\mathcal{O}(10^4)$ terms are considered in the series solution~\cite{Cook:2014cta,Nollert:1999ji}.}}~\cite{leaver85,Nollert:1999ji}. 
%
%
%
%
}
%
%
%
%
\bpar{Polynomial/non-polynomial duality}
The \ccHp{s}, $\ket{u_p}$, may be applied to the \cHp{} problem.
In doing so, a quantitative example of polynomial/non-polynomial duality is encountered, along with related insights into the \qnm{} problem.
\par The \cHp{} problem was stated in \eqn{p22a} as,
\begin{subequations}
    \begin{align}
        \label{p22a-2}
        \mcD{_\xi} \; \ket{y_{pk}}  \; &= \; (\lambda_{pk} \; + \; \mu_{p} \, \xi) \; \ket{y_{pk}} \;,
    \end{align}
    To apply the \ccHp{s}, \eqn{p22a-2} may be restated as
    \begin{align}
        \label{p22a-3}
        \mcL_{\xi}^{(p)} \; \ket{y_{pk}}  \; &= \; \lambda_{pk}  \; \ket{y_{pk}} \;,
    \end{align}
    where 
    \begin{align}
        \mcL_{\xi}^{(p)} \; = \; - \; \mu_{p} \, \xi \; + \; \mcD{_\xi} \;  .
    \end{align}
\end{subequations}
\begin{figure}[t]
    
    
    \begin{tabular}{c} 
        \hspace{0.10cm}\includegraphics[width=0.47\textwidth]{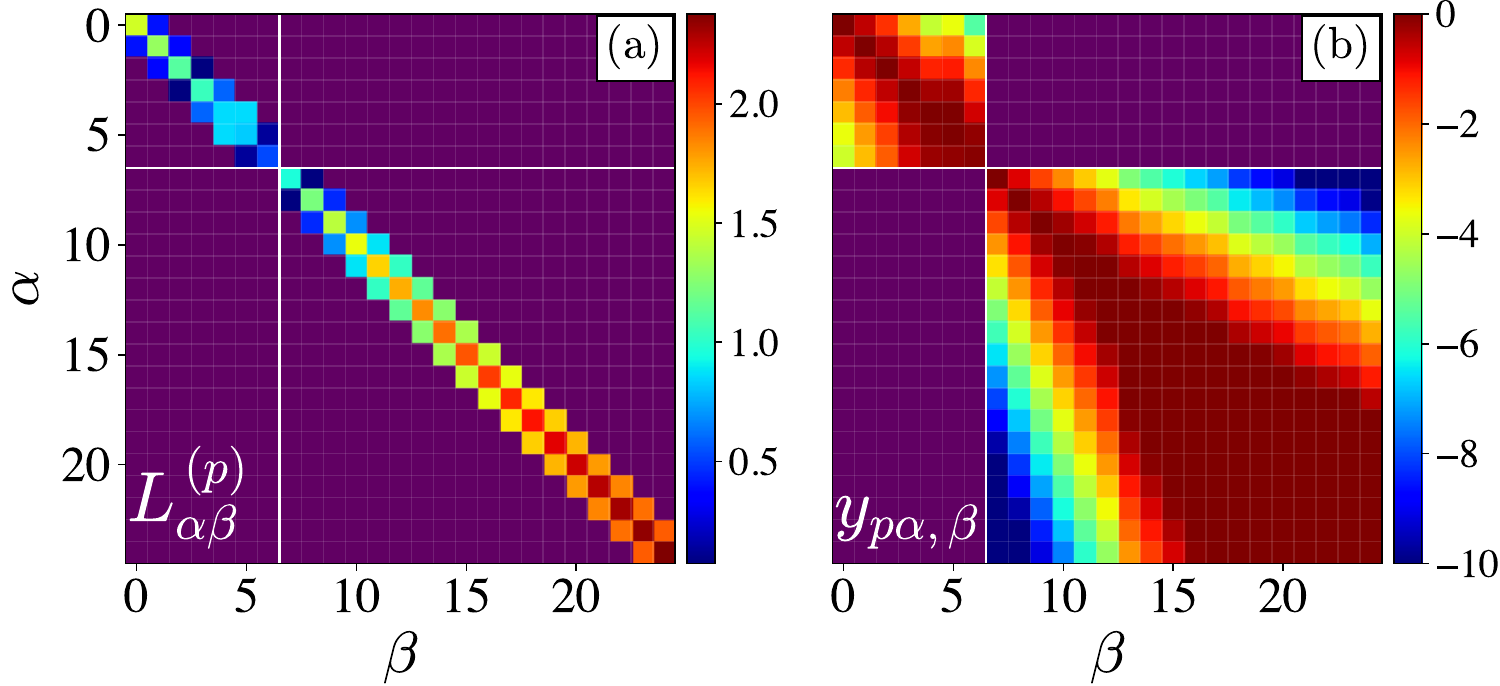} 
        \\
        \hspace{-0.52cm}\includegraphics[width=0.46\textwidth]{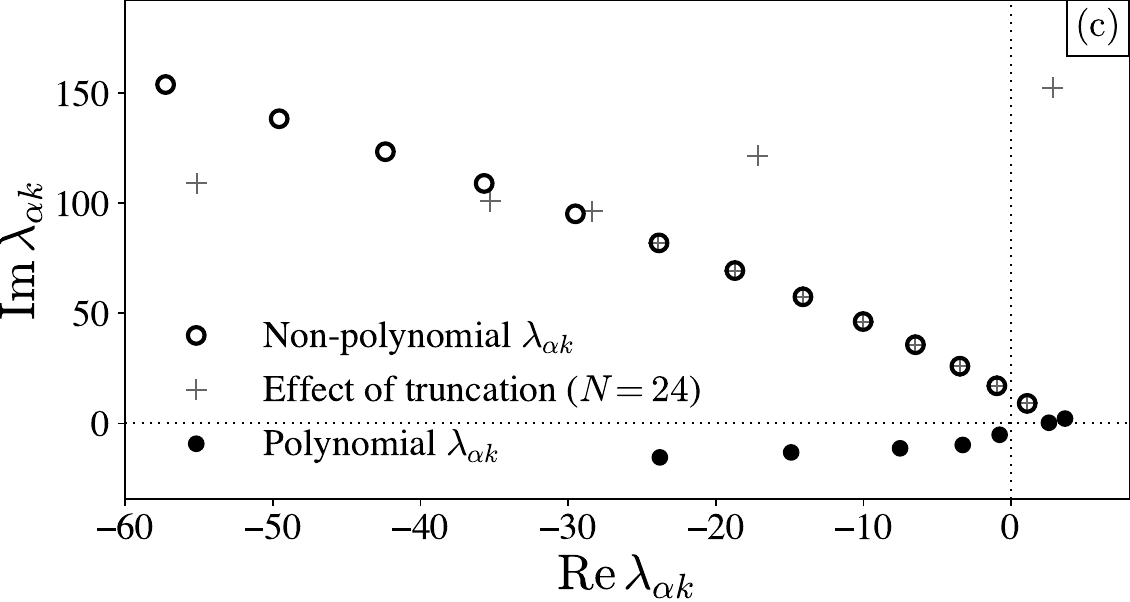}
    \end{tabular}
        
    \caption{ 
        Example of confluent Heun polynomial/non-polynomial duality for polynomial order $p=6$, and physical parameters $(s,\ell,m,n)=(-2,2,2,6)$, $a/M=0.7$, and $M\cw_\lmn=0.4239-1.0954i$. 
        All color bars correspond to the $\log_{10}$ of the absolute values of quantities noted in annotation. 
        Matrix elements of zeo value are colored purple.
        {Panel $(a)$}: Matrix representation of the radial operator~(\ceqn{p53}).
        {Panel $(b)$: Matrix representation of the radial operator's eigenvectors~(\ceqn{p22a-4}).}
        In panels $(a)$ and $(b)$, vertical and horizontal white lines delineate polynomial and non-polynomial sectors. 
        Panel $(c)$: The radial operator's polynomial eigenvalues (filled circles) and the first $13$ non-polynomial eigenvalues (open circles).
      }
    \label{F4}
\end{figure}
By using the \ccHp{s}, \eqn{p22a-3} may be rewritten as 
\begin{align}
    \label{p22a-4}
    \hat{L}^{(p)} \, \vec{y}_{pk} \; = \; \lambda_{pk} \, \vec{y}_{pk} \; ,
\end{align}
where $\hat{L}^{(p)}$ and $\vec{y}_{pk}$, respectively, have elements given by
\begin{subequations}
\begin{align}
    {L}^{(p)}_{\alpha \beta} \; &= \; \brak{u_\alpha}{\mcL_{\xi}^{(p)}|\,u_\beta} \; ,
    \\
    {y}_{pk,\beta} \; &= \; \brak{u_\beta}{y_{pk}} \; .
\end{align}
\end{subequations}
Since $k$ and $\alpha$ are both integers between $0$ and $p$, the sequence of all eigenvectors comprise the rows of a matrix whose elements are ${y}_{p\alpha,\beta}$.
Similarly, the eigenvalues may be labeled as $\lambda_{p\alpha}$ rather than $\lambda_{pk}$.
\smallskip
\par At focus in \fig{F4} are the \cHp{} problem's matrix representation, eigenvectors, and eigenvalues for order-$6$ polynomials.
Physical parameters used are identical to those in \tbl{pspm-examples} for $(\ell,m,n)=(2,2,6)$.
\par In panel $(a)$ of \fig{F4}, ${L}^{(p)}_{\alpha \beta}$ are shown. 
There it may be observed that
\begin{align}
    {L}^{(p)}_{6,7} \; = \; {L}^{(p)}_{7,6} \; = \; 0 \; .
\end{align}
This is consistent with the so-called ``$\Delta_{p+1}$ condition'' which is used to define the \cHp{s} via a finite dimensional matrix eigenvalue problem~(see Eq. 79 of \PaperOne{} as well as Refs.~\cite{NIST:DLMF:ConfHeunPoly,Cook:2014cta}).
There the basic concept is that a power series solution to \eqn{p22a-3} will terminate after $p+1$ terms if the $\Delta_{p+1}$ condition on a related three-term recursion is satisfied. 
\par The $\Delta_{p+1}$ condition is unable to provide insight on the non-polynomial solutions, simply due to fact that these solutions have series expansions of infinite length.
In other words, the non-polynomial solutions satisfy an infinite dimensional matrix eigenvalue problem, rather than a finite dimensional one\footnote{
    In such cases, standard routes to finding solutions include the use of continued fractions (see e.g. Ref.~\cite{leaver85}) or the representation of the problem with an orthonormal basis, such as the \ccHp{} basis introduced here. 
}.
\begin{figure*}[t] 
    
    
    \begin{tabular}{cccc} 
        \hspace{-1cm} 
        \includegraphics[width=0.2972\textwidth]{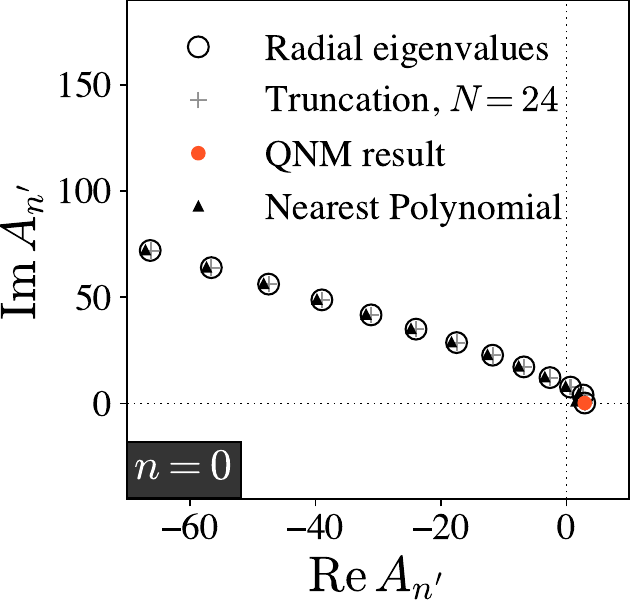} 
        &
        \includegraphics[width=0.2375\textwidth]{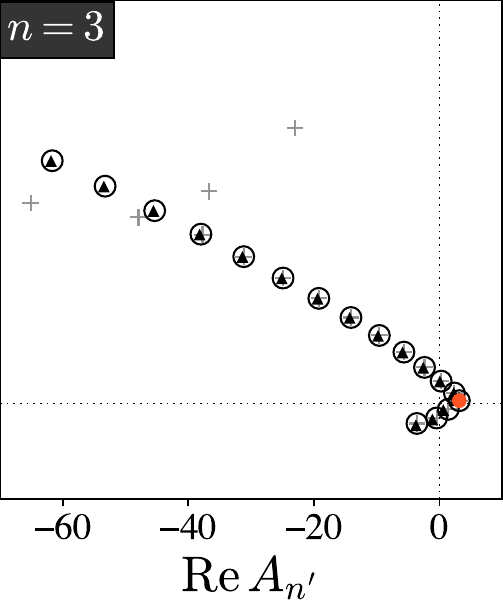} 
        &
        \includegraphics[width=0.2375\textwidth]{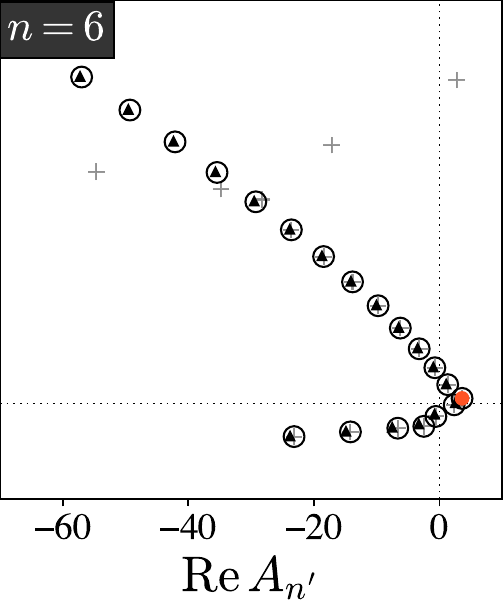} 
        &
        \includegraphics[width=0.2375\textwidth]{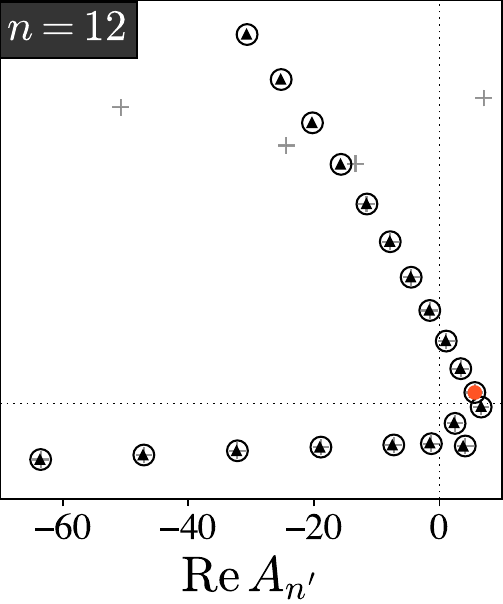} 
    \end{tabular}
        
    \caption{ 
        Radial eigenvalue distributions for select quasinormal mode cases, all with \bh{} dimensionless spin $a/M=0.7$, and angular indices $(\ell,m)=(2,2)$.
        In all panels, radial eigenvalues, $A_{n'}$, are noted with open circles, and the known \qnm{} eigenvalue, $A_{\ell m n}$, is shown as a red filled circle~(c.f.~\cfig{F1}). 
        Left to right: Radial eigenvalues for the respective $n \in \{0,3,6,12\}$ \qnm{} overtones.
        Respective absolute differences between known \qnm{} eigenvalues (Refs.~\cite{positive:2020,Stein:2019mop}) and those in open circles (i.e. $|A_{\ell m n}-A_{n}|$) are $\{2.72\,\times 10^{-9},4.72\,\times 10^{-9},1.24\,\times 10^{-9},6.28\,\times 10^{-9}\}$. 
        \greenoff{Eigenvalues for the \cHp{s} with order $[\tx{Re}\; \psp]$ (\ceqn{near-int}) are shown by black triangles.}
        Center right ($n=6$): See panel $(c)$ of \cfig{F4} for comparison. 
      }
    \label{F5}
\end{figure*}
Unlike the power-series perspective of the $\Delta_{p+1}$ condition, the \ccHp{} representation of the problem naturally applies to finite {and} infinite dimensional problems: 
i.e., the monomials are the basis for power series expansions, but unlike the \ccHp{}, monomials are not an {orthonormal} basis, meaning that standard matrix interpretations do not apply for non-terminating series.
Since the \ccHp{} are orthonormal, the polynomial and non-polynomial sectors are clearly visible in \FigFourA{}.
\par At focus in \FigFourB{} are the eigenvectors; rows labeled in $\alpha$ correspond to vectors $\vec{y}_{p\alpha}$.
There, the polynomial solutions are shown as rows that have zero values after $\beta=6$, as is equivalent to each \cHp{} only being comprised of \ccHp{s} of order $6$ and below.
The non-polynomial solutions are seen as the lower right-hand block, for which values are non-zero only for $\beta>6$, i.e. the non-polynomial solutions do not contain \ccHp{s} with order $p\le 6$.
\par Polynomial and non-polynomial eigenvalues are shown in \FigFourC{} as points in the complex plane.
There, polynomial eigenvalues show a distinctly different distribution than their non-polynomial counterparts. 
If a \qnm{} is approximately polynomial as described in \sec{s3}, then its eigenvalue distribution should be expected to also reflect polynomial/non-polynomial duality.
%
%
\bpar{QNM radial eigenvalue distributions}
It was noted in \sec{s3} that the $(\ell,m,n)=(2,2,6)$ \qnm{} may be approximately polynomial for a \bh{} spin of $a=0.7$~(\ctbl{pspm-examples}). 
In particular, this \qnm{'s} upper pseudo-polynomial order~(\ceqn{pstar}) is 
\begin{align}
    \psp \; &= \; \refpsp \; .
\end{align}
Since the imaginary part of $\psp$ is small, and the real part is close to $6$~(c.f. \ceqn{near-int}), it is reasonable to expect rough agreement between that \qnm{'s} radial eigenvalues, and the order-$6$ \cHp{'s} polynomial \textit{and} non-polynomial eigenvalues~(i.e. $\psp \approx  6$).
\par In \fig{F5}, the radial eigenvalue distributions for four select \qnm{s} are shown. 
There, as $n$ increases, so does the approximately polynomial sector of the eigenvalue distribution. 
\greenoff{The nearest \cHp{} eigenvalues have been added for comparison~(See \ceqn{near-int}). 
For all cases, the qualitative features of the \qnm{} eigenvalue distribution are captured by that of the \cHp{} eigenvalues. 
For example, the $n=6$ \qnm{} separation constant, $A_{226}$, is indeed well approximated by the dominant order-$6$ \cHp{} eigenvalue},
\begin{align}
    \label{cHpA-2}
    A_{226} \; \approx \; \CZero+\COne+\lambda_{66}\; .
\end{align} 
In \eqn{cHpA-2}, \eqn{cHpA} has been restated in the current context, and the parameters, $\CZero$ and $\COne$, have been computed with the relevant  \qnm{} frequency $\cw_{226}$. 
The fractional error between the left and \rhs{s} of \eqn{cHpA-2} is $4.6\%$.
\greenoff{A more thorough explorarion of the \qnm{} solution space is needed to determine whether the qualitative agreement in \fig{F5} is robust. }
\begin{figure*}[t] 
    
    \begin{tabular}{cccc} 
        \hspace{-0.5cm} 
        \includegraphics[width=0.2685\textwidth]{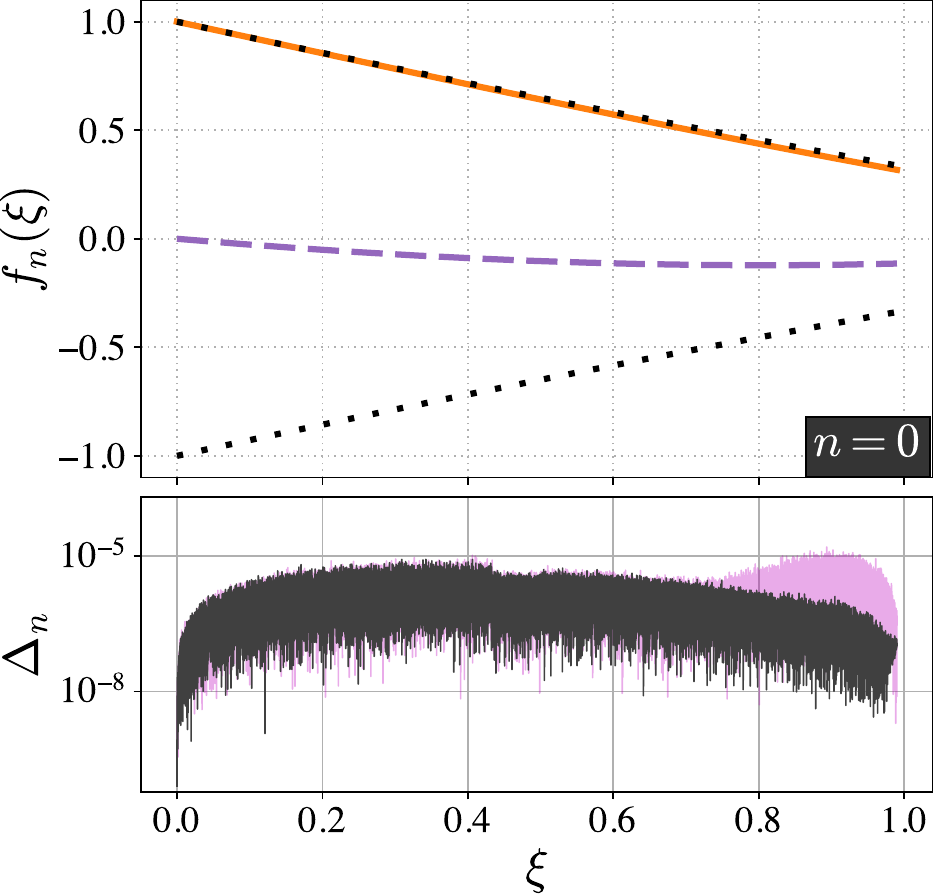} 
        &
        \includegraphics[width=0.230206\textwidth]{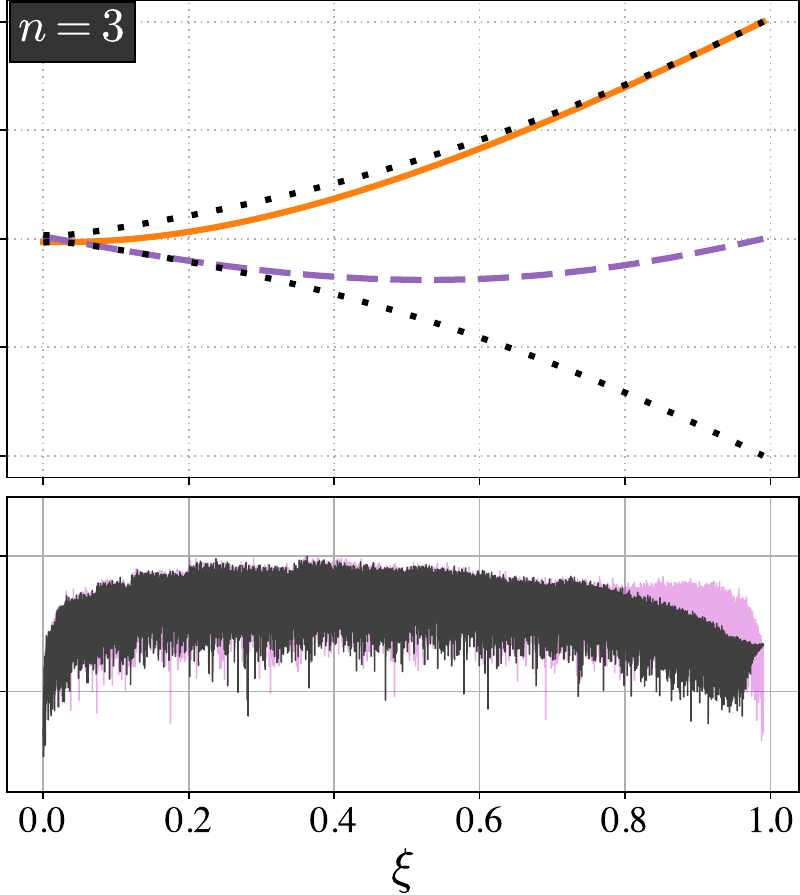} 
        &
        \includegraphics[width=0.230206\textwidth]{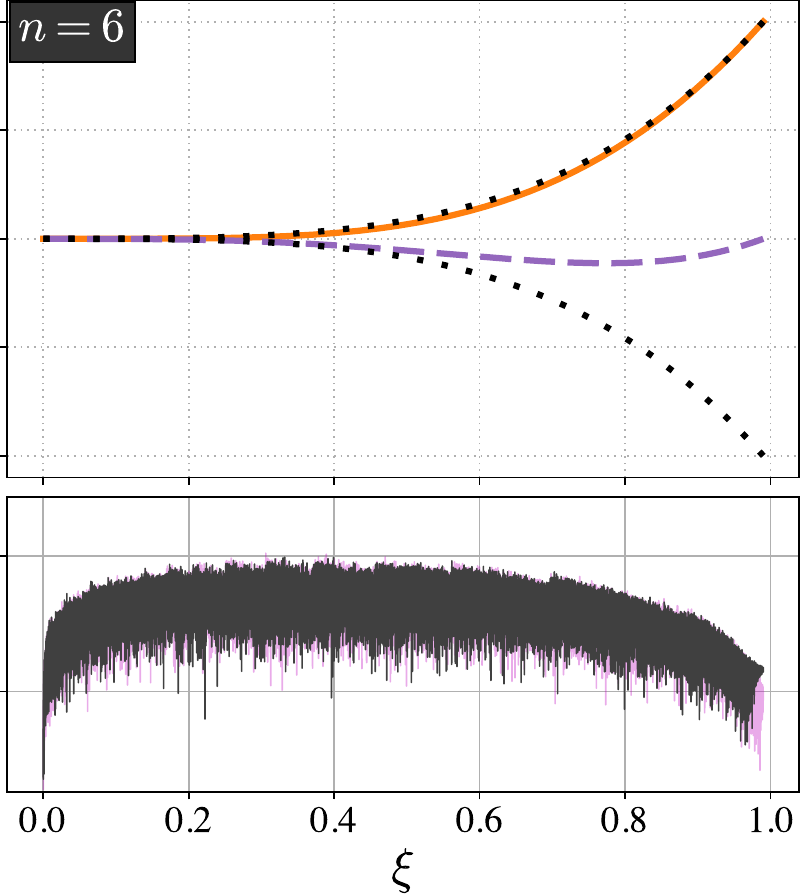} 
        &
        \includegraphics[width=0.230206\textwidth]{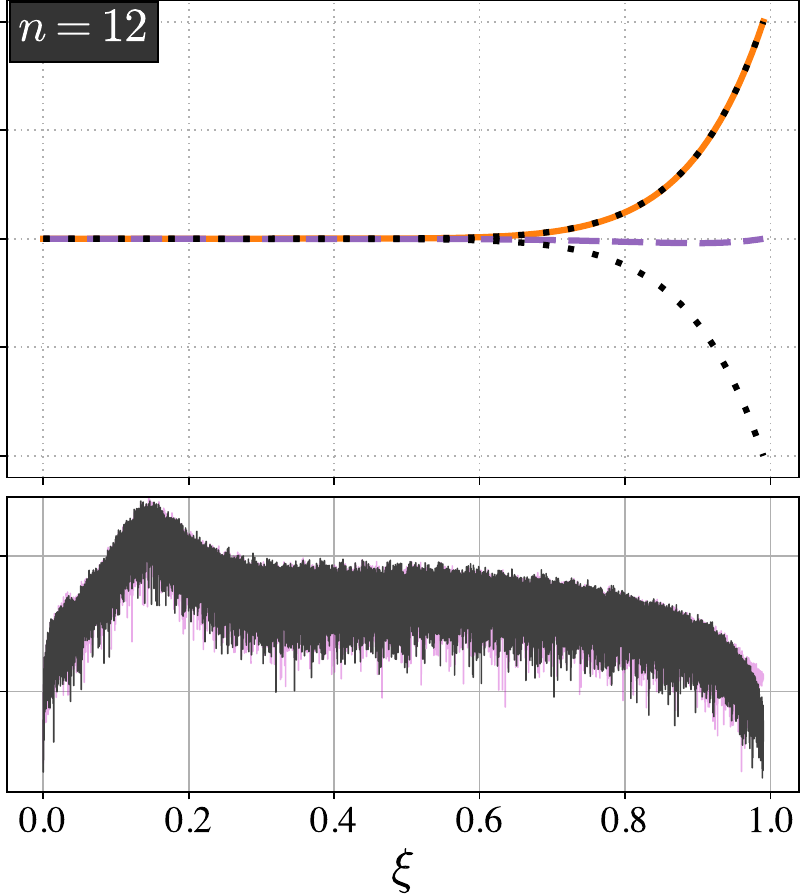} 
    \end{tabular}
        
    \caption{ 
       Radial functions, $f_n(\xi)$, and related floating point error estimates, $\Delta_n$, for the same cases shown in \fig{F5}.
       Top row: The real part of each radial function, $\tx{Re}\,f_n(\xi)$, is denoted with a solid orange line.
       The imaginary part, $\tx{Im}\,f_n(\xi)$, is denoted with a dashed purple line, and the function's positive and negative magnitude, $\pm|f_n(\xi)|$ are denoted with dotted black lines. 
       Bottom row: errors for each radial function, $\Delta_n$~(\ceqn{fn3}). 
       \redoff{In black are errors for $N=24$, where all eigenvectors are consistent with the infinite dimensional problem. In magenta are errors for $N=24$, where eigenvectors due to truncation are included.}
       See \sec{s6} for related discussion.
      }
    \label{F6}
\end{figure*}
%
%
\bpar{Validation of radial functions}
When represented with \ccHp{s}, the \qnm{s'} radial problem may be written as 
\begin{align}
    \label{fn0}
    \mcL{_\xi} \, \ket{f_n} \; &= \; A_n\, \ket{f_n} \; .
\end{align}
In \eqn{fn0}, we have simply rewritten \eqn{p13a} with the overtone label, $n$, explicitly added.
When represented with \ccHp{s}, the radial functions, $f_n(\xi)$, are
\begin{align}
    \label{fn1}
    \ket{f_n} \; = \; \sum_{k=0}^{\infty} \, \ket{u_k}\brak{u_k}{f_n} \;.
\end{align}
In coordinate notation, this is simply 
\begin{align}
    \label{fn2}
    {f_n}(\xi) \; = \; \sum_{k=0}^{\infty} \,  {u_k}(\xi) \, \brak{u_k}{f_n}   \;.
\end{align}
Upon solving the matrix eigenvalue problem~(\ceqn{p45b}), the accuracy of the each $f_n(\xi)$ may be quantified by 
\begin{align}
    \label{fn3}
    |(\mcL_\xi-A_n)f_n(\xi)|/|f_n(\xi)| \; = \; \Delta_{n}(\xi) \; .
\end{align}
In \eqn{fn3}, \eqn{fn0} has been rearranged such that all quantities appear on the \lhs{}.
The absolute value has then been taken to yield a single, real-valued error measure. 
If $f_n(\xi)$ are eigenfunctions, and the derivatives within $\mcL_\xi$ can be evaluated with infinite precision, then $\Delta_{n}(\xi)$ would equal zero for all values of $\xi$.
In practice, $\Delta_{n}(\xi)$ will be very small relative to one, and dominated by numerical noise. 
\par In \fig{F6}, example radial functions are shown along with their errors for the same \qnm{} cases discussed in \fig{F5}.
There, errors have been computed using floating-point precision and spline interpolation for the derivatives~\cite{2020NumPy-Array,positive:2020}.
Despite the result being significantly larger than the calculation's original precision, all cases are observed to have dimensionless errors on the order of $10^{-7}$.
In the top row of panels, radial functions have been scaled such that their largest absolute value is $1$. 
There, it may be observed that as the overtone number increases, the relative value of the radial function at the \bh{} event horizon appears to decrease, possibly implying tension with the physical (asymptotic) boundary conditions, which require $f_n(\xi)$ to be non-zero when $\xi\in\{0,1\}$.
%
\bpar{Schwarzschild and Kerr}
\begin{figure*}[ht] 
    
    \begin{tabular}{cccc} 
        \hspace{-0.5cm} 
        \includegraphics[width= 0.26\textwidth]{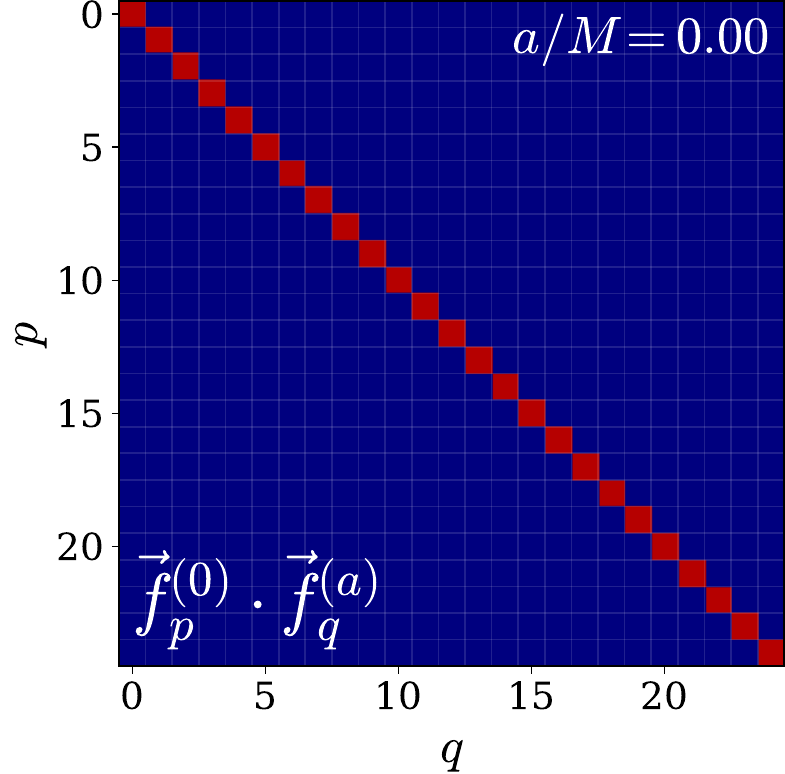}
        &
        \includegraphics[width=0.22\textwidth]{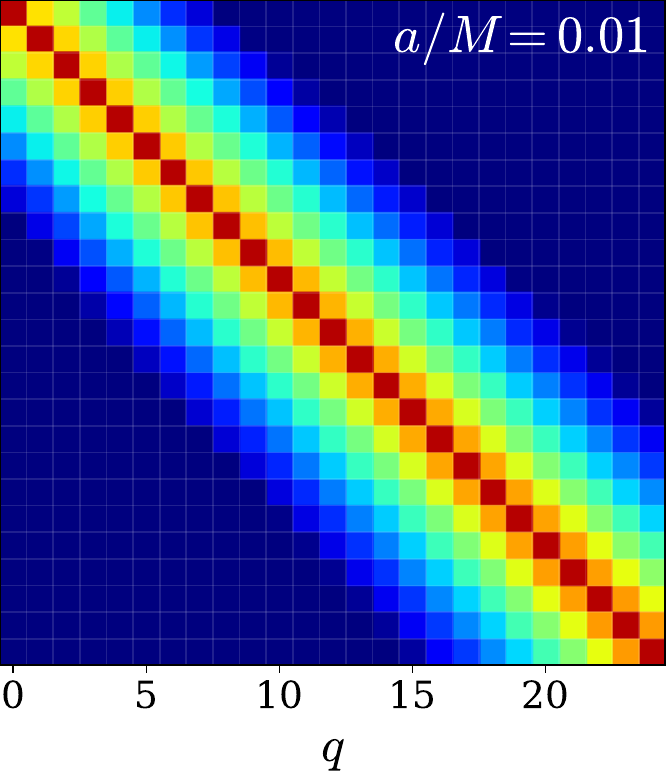}
        &
        \includegraphics[width=0.22\textwidth]{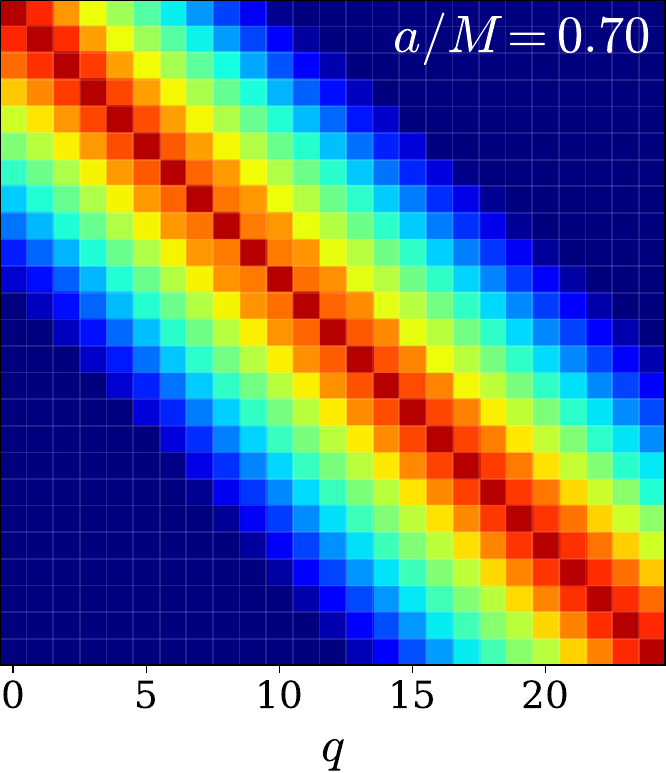}
        &
        \includegraphics[width=0.2615\textwidth]{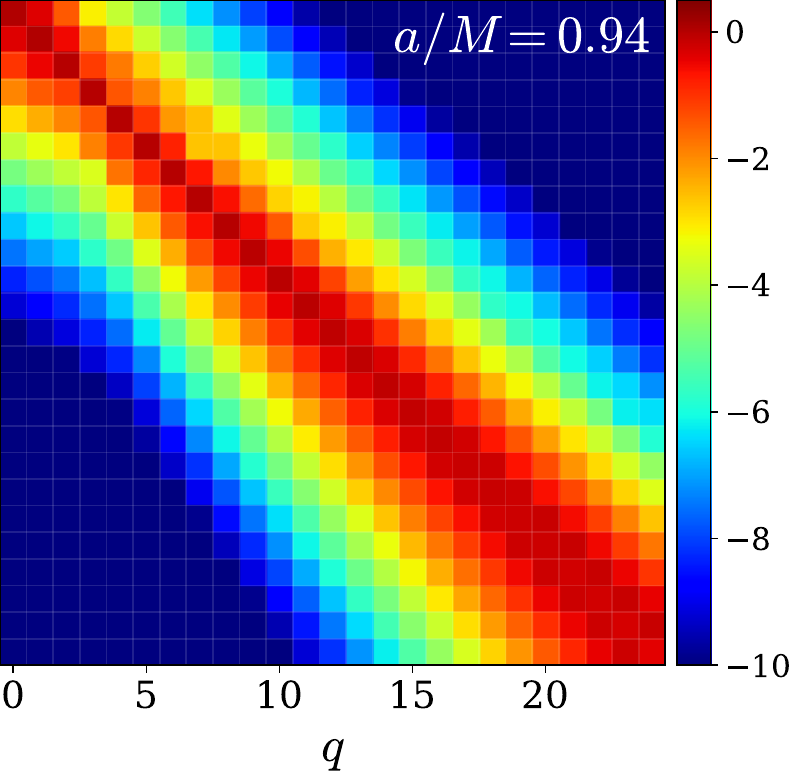} 
    \end{tabular}
        
    \caption{ 
       Mixing matrices between Schwarzschild and Kerr radial functions for four values of \bh{} spin: $a/M\in\{0,0.01,0.70,0.94\}$.
       For each case, the first $24$ radial functions have been computed using \ccHp{s} derived from the $s=-2$, $(\ell,m,n)=(2,2,0)$ \qnm{} frequencies\redoff{; truncation effects have not been included.} 
       \greenoff{Different values of $p$ and $q$ correspond to different values of $n'$ as discussed in \sec{prelims}.}
       {Mixing matrices have been computed according to \eqn{fnProd-d}.}
      }
    \label{F7}
\end{figure*}
Together, the previous examples enable the following observations.
When considered for a single frequency parameter, $\cw$, the radial functions typically have distinct eigenvalues~(\cfig{F5}), and so are linearly independent~\cite{Axler:2015}.
Although the space of eigenfunctions is countably infinite, finite dimensional truncations of this space may be solved for using \ccHp{s}.
In practice, the radial functions will be finitely many and linearly independent, and will therefore constitute a basis.
\par \greenoff{Further, due to the self-adjoint nature of the radial operator \wrt{} the scalar product~(\ceqn{p13}), the space of radial functions is orthogonal~\cite{London:202XP1,lax2002functional,Morse:1955aqj}.}
Since the \ccHp{} are an orthonormal basis, scalar products between different radial functions, $\ket{f_p}$ and $\ket{f_q}$, are equivalent to the simple vector dot product between their vector representations,
\begin{subequations}
    \label{fnProd}
    \begin{align}
        \label{fnProd-a}
        \brak{f_p}{f_q} \; &= \; \sum_{jk} \, \brak{f_p}{u_j}\brak{u_j}{u_k}\brak{u_k}{f_q} \; 
        \\ 
        \label{fnProd-b}
        &= \; \sum_k \, \brak{f_p}{u_k}\brak{u_k}{f_q} \; 
        \\
        \label{fnProd-c}
        &= \; \vec{f}_p \cdot \vec{f}_q \; .
    \end{align}
\end{subequations}
\greenoff{Note that, in \eqn{fnProd}, \ccHp{s} constitute spin-dependent bases. }
Note also that the radial functions may be normalized in the standard way, $\ket{f_p}\leftarrow \ket{f_p}/\sqrt{\brak{f_p}{f_p}}$.
In other words, when computed for a single frequency parameter, the radial functions define an orthonormal basis. 
\par \redoff{A important question in single \pt{} is whether \qnm{s} for one physical case may be used to efficiently represent \qnm{s} in another. 
For example, it is common to represent the Kerr \qnm{s'} angular functions in the basis of those for Schwarzschild~\cite{London:202XP1,London:2020uva,Berti:2014fga,OSullivan:2014ywd,Cook:2014cta,Press:1973zz}.}
Here, we are concerned with whether the same is true for the \qnm{s'} radial functions.
\greenoff{For simplicity, we will consider different \bh{} spins, a single \qnm{} frequency for each, and related families of single-frequency solutions to the radial equation.}
In effect, we ask whether there exists a mixing matrix between sets of radial functions derived from different \bh{} spins\footnote{\greenoff{Note that the invertibility of such a matrix was found to be a prerequisite to the existence of the adjoint-spheroidal harmonics presented in Ref.~\cite{London:2020uva}.}}.
%
%
\par To answer this question, it cannot be avoided that, even for a single \qnm{}, different values of \bh{} spin correspond to different \qnm{} frequencies~\cite{London:2018nxs,leaver85,Berti:2014fga,Berti:2016lat}.
\redoff{When combined with the fact that all forms of \tk{}'s radial equation~(\ceqn{p10}) are frequency dependent, this means that the \qnm{} radial functions for different \bh{} spins are members of {different} scalar product spaces.
While a complete treatment of the question may require the use of a frequency-independent weight function, we remain free to apply \textit{any} scalar product to the problem.
Any choice of scalar product will impact the symmetry properties of $\mcL_\xi$ and the orthogonality properties of the radial functions. }
%
%
\par Let the $p^\tx{th}$ radial function for a given \bh{} spin, $a$, be denoted $\vec{f}_p^{\,(a)}$.
As mentioned previously, if $\vec{f}_p^{\,(a)}$ are defined with \qnm{} frequency $\cw_\lmn$, then $\vec{f}_n^{\,(a)}$ is the \qnm{}'s radial function.
Since each $a$-dependent \ccHp{} basis is orthonormal, \eqn{fnProd} may be interpreted to mean that the scalar product for all $\vec{f}_p^{\,(a)}$ is given by the \rhs{} of \eqn{fnProd-c}.
Since the \rhs{} of \eqn{fnProd-c} has no weight function, we may {superficially} conclude that the standard vector dot product is a frequency-independent scalar product of potential use. 
%
%
%
%
\par Let the scalar product between a Schwarzschild~(i.e. $a=0$) and Kerr radial function be 
\begin{align}
    \label{fnProd-d}
    \vec{f}_p^{(0)} \cdot \vec{f}_q^{(a)} \; = \; \sum_k \, \brak{f_p^{(0)}}{u_k^{(0)}}\brak{u_k^{(a)}}{f_q^{(a)}} \;   .
\end{align}
In \eqn{fnProd-d}, $u_k^{(a)}$ are \ccHp{s} derived from a \bh{} spin of $a$, where $(s,l,m,n)$ are implicitly fixed.
\greenoff{Further, bra-ket operations use the weight defined by the labeled spin value (e.g. $\brak{f_p^{(0)}}{u_k^{(0)}}$ uses \eqn{p14} defined with $a=0$).}
Vectors $\vec{f}_p^{(0)}$ and $\vec{f}_q^{(a)}$ will be considered normalized, making e.g. the set of all $\vec{f}_p^{(0)}$ orthonormal.
\greenoff{Therefore, the truncated set of Schwarzschild radial functions may constitute an orthonormal basis with a well defined projection operator. 
The application of this projection operator to $\vec{f}_{q}^{(a)}$ is equivalent to the decomposition of a Kerr radial function onto the basis of Schwarzschild radial functions. This is}
\begin{align}
    \label{fnProd-e}
    \vec{f}_{q}^{(a)} \; = \; \sum_{p=0} \; \vec{f}_p^{(0)}  \; (\vec{f}_p^{(0)} \cdot  \vec{f}_q^{(a)}) \; .
\end{align}
In \eqn{fnProd-e}, $\vec{f}_p^{(0)} \cdot  \vec{f}_q^{(a)}$ are elements of the mixing matrix between the Schwarzschild and Kerr radial functions. 
In the case of $a=0$, \eqn{fnProd-e} simply reduces to $\vec{f}_{q}^{(a)}=\sum_{q=0}\vec{f}_{p}^{(0)}\delta_{pq}$, signaling a diagonal mixing matrix.
\smallskip
\par In \fig{F7}, the $(s,\ell,m,n)=(-2,2,2,0)$ \qnm{} frequencies have been used to compute Schwarzschild-Kerr mixing matrices for four values of \bh{} spin\footnote{
    The successful reproduction of \fig{F7} requires that one keep in mind the overall sign ambiguity discussed in \sec{s4}:
    Given canonical polynomials, $u_p^{(a)}$ and $u_p^{(a')}$, one can avoid the presence of an incongruous minus sign by using, e.g., the scalar product space of $u_p^{(a)}$ to compute the matrix whose diagonal elements are $\brak{u_p^{(a)}}{u_p^{(a')}}$. 
    Diagonal elements with a negative real part signal that $u_p^{(a')}$ should be scaled by $-1$.
}.
As seen from left to right in \fig{F7}, as \bh{} spin increases, so too does the magnitude of off-diagonal matrix elements.
Nevertheless, the mixing matrix appears to be approximately diagonal throughout the range of spins shown. 
\smallskip
\par This result implies that there exists a one-to-one relationship between Schwarzschild and Kerr radial functions across a wide range of \bh{} spins.
\smallskip
\par Such a one-to-one relationship is known to exist between the Schwarzschild and Kerr angular functions~\cite{London:2020uva,Berti:2014fga}.
These functions are the spin weighted spherical and spheroidal harmonics, respectively~\cite{NP66,Goldberg:1966uu,London:2020uva,OSullivan:2014ywd,Cook:2014cta}. 
It was found in Ref.~\cite{London:2020uva} that the existence of a spherical-to-spheroidal mapping underpins the existence of another set of angular functions called adjoint-spheroidal harmonics, the significance of which may be understood thusly: 
Due to their different \qnm{} frequencies, the \qnm{s'} spheroidal harmonics are not orthogonal with each other; however, they are \textit{bi}orthogonal with the adjoint-spheroidal harmonics.
Therefore the adjoint-spheroidal harmonics signal \qnm{} orthogonality between modes of different $\ell$.
This perspective also defines the sense in which the \qnm{s} are complete in $\ell$~\cite{London:2020uva}.
\par In this context, \fig{F7} implies the potential bi-orthogonality and completeness of the \qnm{s'} radial functions, i.e. the set of all \qnm{} radial functions, where each radial function has frequency $\cw_\lmn$, and where only $n$ is allowed to vary.
However, given the close relationship between the \qnm{s'} radial functions and the \cHp{s}, it is possible that the radial functions are not complete, but rather \textit{overcomplete}~(\sec{s3}).
Indeed, overcompleteness is perhaps most likely, given the known existence of \qnm{} multiplets~\cite{Cook:2014cta,MaassenvandenBrink:2003as}. 
%
%
\section{Discussion and conclusions}
\label{discus}
The present work is based upon a simple premise: to solve \tk{}'s radial equation, rather than looking to special functions from other problems, we may need only look to the equation itself~\cite{London:202XP1,Fiziev:2009}.
The result is a method for computing global solutions to the confluent Heun equation via a single expansion in polynomials that are natural to the problem.
We have introduced such polynomials in \sec{s4}.
In \sec{s5}, we have shown that these polynomials enable the radial equation to be represented as a matrix that is both tridiagonal and symmetric. 
Lastly, in \sec{s6}, we have demonstrated the use of these polynomials in various numerical examples, the last of which may be of particular relevance to the time-dependent perturbations of \bh{s}~\cite{Green:2022htq,Cannizzaro:2023jle,Redondo-Yuste:2023ipg,Sberna:2021eui,Lagos:2022otp}.
We now briefly discuss connections and implications that have not been fully articulated thus far. 
\par As noted previously, one of our goals has been to gain a deeper mathematical understanding of the \qnm{'s} radial problem. 
Consequently, our method for solving \tk{}'s radial equation is particular to the \qnm{s}. 
If our method can be generalized to non-\qnm{} boundary conditions, such as those of a point particle orbiting a \bh{}, then new techniques of analysis and methods of evaluation may result.
Concurrently, it may also be useful to reproduce our results for other formulations of \bh{} perturbation theory, and in more exotic spacetimes~\cite{Mino:1997bx,Berti:2009kk,Hughes:2000pf,Ripley:2022ypi}. 
\par Of the techniques presented here, a number are open to refinement, further exploration, and connection with other recent results. 
For example, we have shown that, in some cases \greenoff{(particularly for large $n$)}, the \qnm{}'s separation constant is well approximated by the eigenvalue of a \cHp{}.
We expect that more analytic understanding is tractable when either representing the problem in \cHp{s}, or their canonical counterparts.
Further, there is good reason to think that more may be learned by a thorough comparison between our results, and those known for Pollaczek-Jacobi polynomials~\cite{Chen:2010,Chen:2019,Min_2023,Yu:10.1063/5.0062949}.
%
%
There is also the possibility of combining the results presented here with recent iterative generalized spectral methods for computing \qnm{} frequencies~\cite{Chung:2023zdq,Blazquez-Salcedo:2023hwg,Chung:2023wkd,ghojogh2023eigenvalue,Zhu:2023mzv}.
\redoff{Regading spectral pollution, it may also be possible to adapt techinques deveoped in quatum meechanics to mitigate the effect without incuring the added computational cost of increased basis size~\cite{Lewin2008arXiv0812.2153L,BOULTON20161,Davies2003math2145D,MartnezAdame2007}.}
\par Lastly, one clear next step is to expand upon the results of \sec{s6} to better understand whether the \qnm{}'s radial functions are complete. 
This would likely require a greater practical understanding of the complex-valued radial coordinate introduced in \PaperOne{}.
Furthermore, this may require an analysis similar to that of Ref.~\cite{London:2020uva} for the spheroidal harmonics.
Intriguingly, if the \qnm{'s} radial functions are complete (in some practical way), then it would follow that any bounded spatial deviation from Kerr might be represented in terms of \qnm{s}.
\section{Acknowledgements}
The authors thank Katy Clough, Stephen Green, Scott Hughes, Laura Sberna and Bernard Whiting for supportive discussions. Lionel London was funded at King's College London by the Royal Society {URF{\textbackslash}R1{\textbackslash}211451}. Michelle Foucoin was funded at King's College London by Royal Society enhancement grant {RF{\textbackslash}ERE{\textbackslash}210040} and by the King's College London department of physics.
\bibliography{references.bib}
\end{document}